\documentclass[conference]{IEEEtran}

\usepackage{float}
\usepackage{svg}
\usepackage{pgfplots}
\usepackage{subcaption}
\usepackage{rotating}
\usetikzlibrary{patterns}

\usepackage{hyperref}
\usepackage{caption}
\let\labelindent\relax
\usepackage{enumitem,kantlipsum}

\hypersetup{
    colorlinks=true,
    linkcolor=blue,
    filecolor=magenta,      
    urlcolor=blue,
}
\DeclareRobustCommand*{\IEEEauthorrefmark}[1]{%
  \raisebox{0pt}[0pt][0pt]{\textsuperscript{\footnotesize\ensuremath{#1}}}}
  
\usepackage[misc]{ifsym}
\newcommand{\corrAuthor}{$^{\textrm{\Letter}}$}

\begin{document}

\title{
Assessing the Use Cases of Persistent Memory in High-Performance Scientific Computing}


 \author{
 \authorblockN{Yehonatan Fridman\authorrefmark{1,2}, 
 Yaniv Snir\authorrefmark{1,3}, 
 Matan Rusanovsky\authorrefmark{1,2}, 
 Kfir Zvi\authorrefmark{1,2}, 
 Harel Levin\authorrefmark{4},\\
 Danny Hendler\authorrefmark{1}, 
 Hagit Attiya\authorrefmark{5}, 
 Gal Oren\authorrefmark{4,5} \corrAuthor}\\
 \authorblockA{\authorrefmark{1} Department of Computer Science, Ben-Gurion University of the Negev}
 \authorblockA{\authorrefmark{2} Israel Atomic Energy Commission}
  \authorblockA{\authorrefmark{3} Google}
  \authorblockA{\authorrefmark{4} Scientific Computing Center, Nuclear Research Center -- Negev}
 \authorblockA{\authorrefmark{5} Department of Computer Science, Technion -- Israel Institute of Technology}\\
 {\tt\small \{fridyeh, yanivsn, matanru, zvikf\}@post.bgu.ac.il, harellevin@nrcn.org.il,}\\
{\tt\small hendlerd@cs.bgu.ac.il, \{hagit, galoren\}@cs.technion.ac.il}}

\maketitle

\begin{abstract}
As the High Performance Computing (HPC) world moves towards the Exa-Scale era, huge amounts of data should be analyzed, manipulated and stored. In the traditional storage/memory hierarchy, each compute node retains its data objects in its local volatile DRAM. Whenever the DRAM’s capacity becomes insufficient for storing this data, the computation should either be distributed between several compute nodes, or some portion of these data objects must be stored in a non-volatile block device such as a hard disk drive (HDD) or an SSD storage device. These standard block devices offer large and relatively cheap non-volatile storage, but their access times are orders-of-magnitude slower than those of DRAM.
Optane™ DataCenter Persistent Memory Module (DCPMM)~\cite{liu2021survey}, a new technology introduced by Intel, provides non-volatile memory that can be plugged into standard memory bus slots (DDR DIMMs) and therefore be accessed much faster than standard storage devices.\\
In this work, we present and analyze the results of a comprehensive performance assessment of several ways in which DCPMM can 1) replace standard storage devices, and 2) replace or augment DRAM for improving the performance of HPC scientific computations. To achieve this goal, we have configured an HPC system such that DCPMM can service I/O operations of scientific applications, replace standard storage devices and file systems (specifically for diagnostics and checkpoint-restarting), and serve for expanding applications' main memory. We focus on keeping the scientific codes with as few changes as possible, while allowing them to access the NVM transparently as if they access persistent storage. Our results show that DCPMM allows scientific applications to fully utilize nodes’ locality by providing them with sufficiently-large main memory. Moreover, it can also be used for providing a high-performance replacement for 
persistent storage.
Thus, the usage of DCPMM has the potential of replacing standard HDD and SSD storage devices in HPC architectures and enabling a more efficient platform for modern supercomputing applications. \\The source code used by this work, as well as the benchmarks and other relevant sources, are available at: \url{https://github.com/Scientific-Computing-Lab-NRCN/StoringStorage}.

\end{abstract}


\maketitle


\section{Introduction}

Modern scientific research and advanced industries make massive usage of \emph{High Performance Computing} systems.
This makes these systems a critical resource in both private and public sectors, 
supporting compute-intensive calculations that are otherwise impossible to perform~\cite{hager2010introduction}. 
Indeed, many recent achievements in Physics, Chemistry, Engineering, Biology, Medicine, Computer Science and more, rely on computations performed by HPC systems~\cite{kaufmann1992supercomputing}. 
As the demand for larger calculations and simulations increases, 
huge amounts of data should be analyzed, manipulated and stored~\cite{snir2014addressing}. 

In the traditional memory management approach, each compute node stores its data objects in its local volatile DRAM~\cite{jacob2010memory}. 
Whenever the memory requirements of the application exceed the node's DRAM capacity, the computation should either be 1) distributed, along with the node's data objects, between several compute nodes, or 2) accommodated by the node by relying on the operating system's virtual memory support, for storing portions of the node's data objects in storage devices attached to it using paging~\cite{jacob2010memory}. Unfortunately, the gap between the speeds of DRAM and standard block storage devices is about three orders of magnitude~\cite{luttgau2018survey}. As computations grow in scale, especially in the exa-scale era, this performance gap becomes intolerable~\cite{harrod2012journey}. Moving data between DRAM and a block device is costly in terms of application runtime~\cite{jacob2010memory}. On the other hand, DRAM is much more expensive than standard storage devices~\cite{luttgau2018survey} and its power consumption is higher, due to the cost of keeping it fresh and alive~\cite{fan2001memory}. Thus, building cost-effective systems with large memory and storage spaces that are resilient, fast and power-efficient is a major challenge faced by the HPC community~\cite{oren2016memory,bergman2008exascale}. 

Another challenge is that as HPC architectures become more complex, 
with larger numbers of compute nodes, the fault rates of large-scale computations increase~\cite{snir2014addressing}.
At the same time, the runtime penalty incurred by each application crash becomes more and more expensive. Consequently, in addition to the constant improvements in the sustainability of HPC clusters, the resilience of scientific applications becomes an important requirement. In traditional HPC memory arrangements, whereas most of the frequently-accessed data is stored in volatile DRAM, various techniques are used to save important data to persistent storage so that it will be available for recovery after an application or system failure~\cite{moody2010design,wang2010hybrid}. 
The most common way of persisting such important data is by storing \textit{check-points} along the program's execution. The frequency of check-pointing is determined according to the probability of crashes in the system and the nature of the algorithm. However, as mentioned above, moving data from the DRAM to storage takes a significant amount of time. 

A new technology introduced by Intel, called Optane™ DataCenter Persistent Memory Module (DCPMM), provides non-volatile memory (NVM) that can be accessed much faster than standard storage devices and is plugged to standard memory bus slots (DDR DIMM). Moreover, unlike contemporary storage devices that can only be accessed in block granularity, Optane™ DCPMM memory is byte addressable and can thus be accessed by applications using regular loads and stores. Due to these advantages, Optane™ DCPMM offers a new tier in the memory-storage hierarchy, which fits perfectly in the gap between memory (registers, cache and DRAM) and storage (SSDs and HDDs). 

\subsection*{Contribution}
\noindent In this work, we present a comprehensive performance assessment of storageless persistent-memory supercomputing configurations. We test several such configurations, in which Intel Optane™ DCPMM and its novel utilities 1) fill the gap between standard storage devices and file systems (especially for diagnostics and checkpoint-restarting), and 2) serve as an expansion of the memory hierarchy that can be used directly by applications, partially replacing the need for expensive paging. 
Unlike previous research in this field, 
we focus on scientific benchmarks that are often used for evaluating the performance of parallel supercomputers. We experiment with representative memory and storage usage scenarios scientific applications that run on HPC systems, in order to assess the extent to which NVM can increase productivity for scientific computing. Specifically, we focus on the following three NVM use cases: (1) Using NVM to expand a node's main memory; (2) Using NVM as a fast local storage area; (3) Using a combination of NVM and DRAM for implementing a fast internal and external checkpoint-restart mechanism. 

We evaluate the potential benefits of DCPMM for HPC scientific computing by using widely-used parallel benchmarks, such as NPB and PolyBench (see Section~\ref{Chapter:Benchmarks}). 

Our results show that Optane™ DCPMM allows scientific applications to better utilize nodes’ locality.
It is able to greatly enlarge a node's main memory and can be used also for storing 
high performance file systems. In addition, it can be used for enhanced checkpoint-restart support. Viewed collectively, these results indicate that DCPMM opens up opportunities for new exa-scale NVM-supported supercomputers. The reliance of these new HPC architectures on standard storage devices (HDDs and SSDs) for scientific computing can be greatly reduced, allowing better performance.

\subsection*{Related Work}
\noindent Optane™ DCPMM~\cite{liu2021survey} is a relatively new hardware, 
still not fully present in any supercomputer worldwide.
Thus, it was subjected so far only to limited examination on scientific workloads.
Deployment of Optane™ DCPMM for high-performance scientific applications was evaluated~\cite{weiland2019early, wu2020lessons, weiland2020usage}, 
measuring outright performance, efficiency and usability for both its Memory and App Direct modes. 
However, these works do not assess the feasibility of 
eliminating the intrinsic need for standard storage altogether, including for local and distributed diagnostics and checkpoint-restart flows, as a complete solution. 

Extending node main memory was tested in several benchmarks and production legacy codes~\cite{peng2020demystifying, wu2021runtime, christgau2020leveraging, ren2021optimizing, ren2021optimizing, malinowski2019multi}, measuring the extent of performance degradation for scientific workflows in comparison to DRAM-only usage. 
Previous results for local and distributed persistent memory file systems with NVM are also promising for several real-world applications, exemplifying significant improvements over current state-of-the-art NVMe SSD storage. However, most of the benchmarks employed~\cite{garg2020need, zhu2021empirical, bother2021drop, soumagne2021accelerating, lopez2021exploring} are not scientific workflows or are not based on common stressors for scientific patterns in supercomputing such as BTIO \cite{wong2003parallel} with all of its variants under MPI.
Furthermore, these works do not compare local and distributed storage over RDMA, 
and they also do not compare to common storage devices such as HDD or SATA-SSD,
which are still very common in large-scale distributed storage systems.
The same holds for previous evaluations of the use case of failure recovery, 
either internal or external~\cite{ren2019easycrash, zvi2021optimized, ren2020exploring}, 
via diagnostic to DCPMM and restart from DRAM.

\begin{figure*}[ht!]
\begin{subfigure}{0.3\textwidth}
\subcaptionbox{Experimental environment specifications.
\label{specs}} 
{\scalebox{.8}{
\begin{tabular}{ |c|c| } 
 \hline
 \#Sockets & 2\\ 
 \hline
 CPU Spec. & 10 Cores $\times$ Intel(R) Xeon(R) Gold  \\ & 5215 CPU  @ 2.50GHz (per socket)  \\ 
 \hline
 L1 Cache & 32KB i-Cache \\ & 32KB d-Cache (per core)  \\ 
 \hline
 L2 Cache & 1024KB (per core)  \\ 
 \hline
 L3 Cache & 14080KB (shared, per socket)  \\ 
 \hline
 DRAM Spec. & 16GB DDR4 DRAM 2933 MT/s \\ 
 \hline
 Total DRAM & 192GB [(2 sockets)$\times$ \\ & (6 channels)$\times$16GB] \\ 
 \hline
 NVM Spec. & 256GB Intel Optane™ DCPMM \\ & 2666 MT/s Apache Pass  \\ 
 \hline
 Total NVM & 1024GB [(2 sockets)$\times$ \\ & (2 channels)$\times$256GB]  \\ 
 \hline
 SSD Spec. & 240GB 6GB/s Intel SATA 2.5” \\ & SSD  \\ 
 \hline
 Network & 56Gb/s Mellanox Infiniband FDR \\
 \hline
 Hyper-Threading & disabled \\
 \hline
 OS & Linux CentOS 7.9, Kernel 4.13.0 \\
 \hline
\end{tabular}
}}
\end{subfigure}
\hspace{1.9cm}
\begin{subfigure}{1\textwidth}

  \subcaptionbox{DCPMM population configuration of a node, with two sockets (SO) connected via an interconnect, used in our experimental environment. Each CPU unit has two memory controllers (MC), each providing three memory channels (CH) and each memory channel contains two DIMM slots. D (in red) denotes DCPMMs, R (in blue) denotes DRAMs (i.e RDIMMs), and white denotes vacancy. Red slots refer to 256GB Optane™ DCPMMs, and blue slots refer to 16GB DDR4 RAM DIMMs.
  \label{population}} {\scalebox{.58}{
  \includegraphics[width=\textwidth]{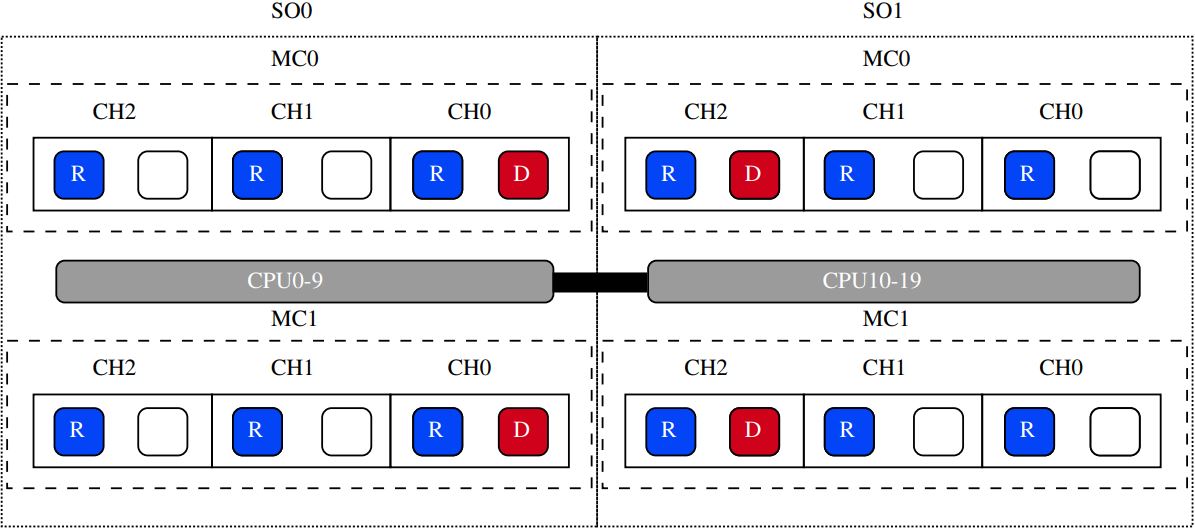}
  }}
\end{subfigure}
\vspace{-0.2cm}
\caption{Experimental environment specifications and DCPMM population.}
\vspace{-0.5cm}
\end{figure*}


\subsection*{Organization}
\noindent The rest of the paper is organized as follows. 
Section~\ref{Chapter:NVRAM} presents the Optane™ DCPMM and the hardware setting we used for this work. In Sections \ref{Chapter:Benchmarks} and \ref{Chapter:ExtendedVirtualMemory}, we describe the well-known NPB and PolyBench scientific application benchmarks and use them for evaluating the performance gained by using NVM as a main memory expansion in comparison to using paging.  
In Section~\ref{Chapter:DistributedStorage}, we evaluate the usage of Optane™ DCPMM as a 
persistent memory storage, using representative local Persistent Memory File Systems (PMFSs). We compare the performance of this solution with that of local SATA SSD using the BTIO specialized benchmark for I/O. 
Finally, Section~\ref{Chapter:CheckpointRestartStorage} presents Optane™ DCPMM as a fast storage for internal (SCR) and external (DMTCP) Checkpoint/Restart storage over local PMFS using the BT solver (from the NPB suite) and compare its performance to that of local SATA SSD. We conclude, in Section~\ref{Chapter:Conclusions}, with a review of Optane™ DCPMM's current functionalities and future potential for scientific computations in the new exa-scale supercomputing era.


\section{Optane™ DCPMM and Hardware Setting}
\label{Chapter:NVRAM}

Traditional architectures~\cite{mcanlis1984data} use non-volatile memory 
only as Double Data Rate (DDR) Dual In-line Memory Modules (DIMMs), 
accessed through standard DDR cards using an ancillary battery.
Upon a power failure, the data is washed to a non-volatile disk.
Intel Optane™ DCPMM uses the same memory connection as DDR DIMMs, and therefore Xeon™ processors can use it as a \emph{standard byte-addressable memory}, in addition to using it as non-volatile memory. Each DCPMM can be configured in one of two modes:

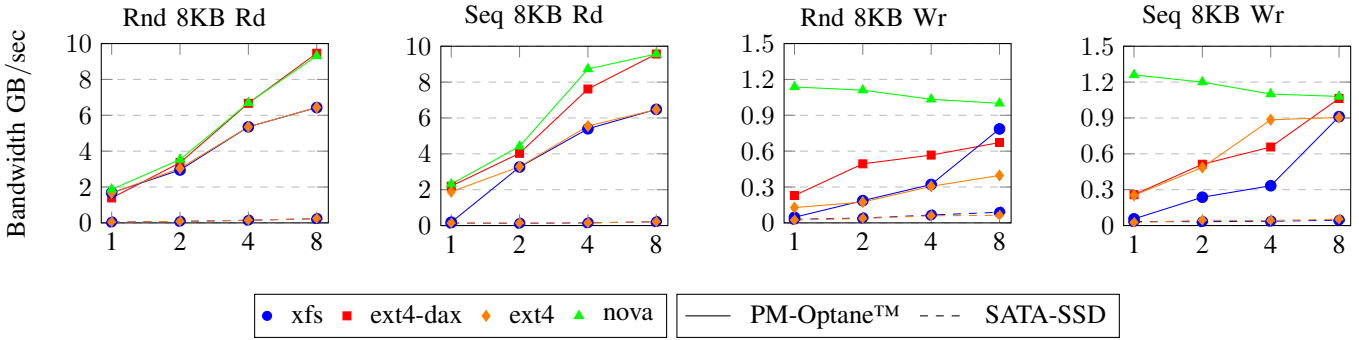
\begin{figure*}[ht!]
\begin{subfigure}{0.01\textwidth}
\begin{turn}{90} 
Bandwidth GB$/$sec
\end{turn} 
\end{subfigure}%
~
\begin{subfigure}[t!]{0.24\textwidth}
\centering
Rnd 8KB Rd
\begin{tikzpicture}
\begin{axis}[
    xmin=0.9, xmax=9,
    ymin=0, ymax=10,
    xtick= {1,2,4,8},
    xticklabels={1,2,4,8,10},
    ytick={0,2,4,6,8,10},
    ymajorgrids=true,
    grid style=dashed,
    width=4.6cm,
    xmode=log,
]

\addplot[
    color=red,
    mark=square*,
    mark size= 1.5pt,
    ]
    coordinates {
    (1,1.39)(2,3.357)(4,6.669)(8,9.46)
    };
\addplot[
    color=blue,
    mark=*,
    dashed,
    ]
    coordinates {
    (1,0.043)(2,0.0845)(4,0.146)(8,0.237)
    };
\addplot[
    color=blue,
    mark=*,
    ]
    coordinates {
    (1,1.678)(2,2.95)(4,5.355)(8,6.430)
    };
\addplot[
    color=orange,
    mark=diamond*,
    ]
    coordinates {
    (1,1.667)(2,3.042)(4,5.355)(8,6.439)
    };
\addplot[
    color=orange,
    mark=diamond*,
    dashed,
    ]
    coordinates {
    (1,0.044)(2,0.082)(4,0.147)(8,0.238)
    };
\addplot[
    color=green,
    mark=triangle*,
    ]
    coordinates {
    (1,1.87)(2,3.53)(4,6.69)(8,9.31)
    };
\addplot[
    color=purple,
    mark=x,
    ]
    coordinates {
    };
\end{axis}

\end{tikzpicture}
\end{subfigure}%
~
\begin{subfigure}[t!]{0.24\textwidth}
\centering
Seq 8KB Rd
\begin{tikzpicture}
\begin{axis}[
    xmin=0.9, xmax=9,
    ymin=0, ymax=10,
    xtick= {1,2,4,8},
    xticklabels={1,2,4,8},
    ytick={0,2,4,6,8,10},
    ymajorgrids=true,
    grid style=dashed,
    width=4.6cm,
    xmode=log,
]

\addplot[
    color=red,
    mark=square*,
    mark size= 1.5pt,
    ]
    coordinates {
    (1,2.2)(2,4.03)(4,7.615)(8,9.566)
    };
    \addplot[
    color=blue,
    mark=*,
    dashed,
    ]
    coordinates {
    (1,0.137)(2,0.123)(4,0.150)(8,0.215)
    };
    \addplot[
    color=blue,
    mark=*,
    ]
    coordinates {
    (1,0.1858)(2,3.264)(4,5.396)(8,6.478)
    };
    \addplot[
    color=orange,
    mark=diamond*,
    ]
    coordinates {
    (1,1.884)(2,3.284)(4,5.549)(8,6.468)
    };
    \addplot[
    color=orange,
    mark=diamond*,
    dashed,
    ]
    coordinates {
    (1,0.137)(2,0.140)(4,0.147)(8,0.211)
    };
       \addplot[
    color=green,
    mark=triangle*,
    ]
    coordinates {
    (1,2.31)(2,4.42)(4,8.73)(8,9.57)
    };
    \addplot[
    color=purple,
    mark=x,
    ]
    coordinates {
    };
\end{axis}
\end{tikzpicture}
\end{subfigure}%
~
\begin{subfigure}[t!]{0.24\textwidth}
\centering
Rnd 8KB Wr
\begin{tikzpicture}
\begin{axis}[
    xmin=0.9, xmax=9,
    ymin=0, ymax=1.5,
    xtick= {1,2,4,8},
    xticklabels={1,2,4,8},
    ytick={0,0.3,0.6,0.9,1.2,1.5},
    ymajorgrids=true,
    grid style=dashed,
    width=4.6cm,
    xmode=log,
]

\addplot[
    color=red,
    mark=square*,
    mark size= 1.5pt,
    ]
    coordinates {
    (1,0.229)(2,0.494)(4,0.567)(8,0.672)
    };
    \addplot[
    color=blue,
    mark=*,
    dashed,
    ]
    coordinates {
    (1,0.030)(2,0.0404)(4,0.0654)(8,0.0891)
    };
    \addplot[
    color=blue,
    mark=*,
    ]
    coordinates {
    (1,0.047)(2,0.186)(4,0.321)(8,0.786)
    };
     \addplot[
    color=orange,
    mark=diamond*,
    ]
    coordinates {
    (1,0.128)(2,0.174)(4,0.306)(8,0.397)
    };
     \addplot[
    color=orange,
    mark=diamond*,
    dashed,
    ]
    coordinates {
    (1,0.0236)(2,0.037)(4,0.0594)(8,0.0627)
    };
     \addplot[
    color=green,
    mark=triangle*,
    ]
    coordinates {
    (1,1.137)(2,1.11)(4,1.034)(8,1)
    };
    \addplot[
    color=purple,
    mark=x,
    ]
    coordinates {
    };
\end{axis}
\end{tikzpicture}
\end{subfigure}%
~
\begin{subfigure}[t!]{0.24\textwidth}
\centering
Seq 8KB Wr
\begin{tikzpicture}
\begin{axis}[
    xmin=0.9, xmax=9,
    ymin=0, ymax=1.5,
    xtick= {1,2,4,8},
    xticklabels={1,2,4,8},
    ytick={0,0.3,0.6,0.9,1.2,1.5},
    ymajorgrids=true,
    grid style=dashed,
    width=4.6cm,
    xmode=log,
]
\addplot[
    color=red,
    mark=square*,
    mark size= 1.5pt,
    ]
    coordinates {
    (1,0.258)(2,0.513)(4,0.657)(8,1.063)
    };
    \addplot[
    color=blue,
    mark=*,
    dashed,
    ]
    coordinates {
    (1,0.0336)(2,0.0325)(4,0.0363)(8,0.0451)
    };
    \addplot[
    color=blue,
    mark=*,
    ]
    coordinates {
    (1,0.056)(2,0.237)(4,0.333)(8,0.911)
    };
    \addplot[
    color=orange,
    mark=diamond*,
    dashed,
    ]
    coordinates {
    (1,0.0255)(2,0.0438)(4,0.0423)(8,0.0531)
    };
    \addplot[
    color=orange,
    mark=diamond*,
    ]
    coordinates {
    (1,0.253)(2,0.486)(4,0.885)(8,0.904)
    };
    \addplot[
    color=green,
    mark=triangle*,
    ]
    coordinates {
    (1,1.26)(2,1.2)(4,1.1)(8,1.08)
    };
    \addplot[
    color=purple,
    mark=x,
    ]
    coordinates {
    };
\end{axis}
\end{tikzpicture}
\end{subfigure}%
\vskip\baselineskip
~
\centering
\newenvironment{customlegend}[1][]{%
    \begingroup
    \csname pgfplots@init@cleared@structures\endcsname
    \pgfplotsset{#1}%
}{%
    \csname pgfplots@createlegend\endcsname
    \endgroup
}%
\def\addlegendimage{\csname pgfplots@addlegendimage\endcsname}

\begin{tikzpicture}
    \begin{customlegend}[legend columns=-1, legend style={column sep=1.2ex}, legend entries={xfs,ext4-dax,ext4,nova}]
    \addlegendimage{blue,mark=*,only marks,mark size=1.7pt}
    \addlegendimage{red,mark=square*,only marks,mark size=1.5pt}
    \addlegendimage{orange,mark=diamond*,only marks,mark size=2pt}
    \addlegendimage{green,mark=triangle*,only marks,mark size=2pt}
    \end{customlegend}
\end{tikzpicture}
\begin{tikzpicture}
    \begin{customlegend}[legend columns=-1, legend style={column sep=1.2ex}, legend entries={PM-Optane™,SATA-SSD}]
    \addlegendimage{black,mark options=solid}
    \addlegendimage{black,dashed,mark options=solid}
    \end{customlegend}
\end{tikzpicture}
    \caption{FIO benchmark results, tested on the Optane™ DCPMM and SATA-SSD of our experimental environment.}
    \vspace{-0.5cm}
    \label{FIO}
\end{figure*}

\begin{enumerate}[wide, labelwidth=!, labelindent=0pt]
    \item \emph{Memory Mode} (also called \emph{Memory-side Cache}, \emph{2-Level-Memory} or \emph{IMDT}): This mode uses the DCPMM as a pool of volatile memory. 
    When DCPMM is configured in this way, the system sees it as its exclusive memory 
    and the DDR serves as the \textit{cache} for the Non-volatile DIMM (NVDIMM).
    This approach is beneficial when a relatively cheap memory expansion is required, because NVDIMM pieces are cheaper than standard DDR pieces and have a much larger volume. A significant advantage of this mode is that it is very easy to adapt the DCPMM to the system (almost "Plug and Play"). 
    However, this mode has several disadvantages in using non-uniform memory
    access (NUMA) codes when running different applications on the cores, 
    since localization is not perfect and DDR cannot really be used as sophisticated cache~\cite{peng2019system}.
    In addition, performance does not scale in this mode as well as with DDR,
    although when the application does not produce many page faults, 
    performance is expected to be close to that of standard RAM~\cite{peng2019system}.

    \item \emph{App Direct} (also called \emph{1-Level-Memory}): 
    This mode uses the persistent memory (PM) directly from the application,
    so that the DDR is used as volatile cache memory while the NVDIMM is used as RAM. 
    The NVDIMM is directly connected to the CPU (via the Memory Bus), 
    and can thus be accessed significantly faster than standard storage. 
    The difference between NVDIMM and NVMe SSD devices (e.g. Intel Optane™ SSD) is that the latter are connected via the PCIe, which adds additional overhead. 
    Moreover, in this mode, \emph{block I/O} can be used as with standard storage, but much faster since the connection is done via the memory bus. 
    However, in order to use the NVDIMM as NVM and access it at the \emph{byte-level} using loads/stores, it has to be mapped to virtual memory via Direct Access (DAX) from the application. 
    The drawback is that direct access requires extra programming effort. 
    In NUMA architectures, Device DAX can be used in App Direct mode to gain access to the images of both NUMA nodes (NUMA node 0 and NUMA node 1) so that the application is able to choose which memory to use for normal allocations. In this work, we chose not to use this feature in order to keep the configuration of the nodes as close to the standard as possible.
 \end{enumerate}
 
The \emph{Persistent Memory Development Kit} (PMDK)~\cite{Scargall2020} is a collection of libraries and tools created by Intel for application developers and system administrators to simplify access and management of persistent memory devices. It consists of volatile libraries (e.g. libmemkind and libvmem) that provide control of memory characteristics, as well as of persistent libraries (e.g. libpmem and libpmemobj), which help applications maintain the consistency of data structures in the presence of failures. These libraries provide new semantics that require extensive application modifications, and in some cases even necessitates building the applications from the ground up \cite{Scargall2020}. In this work, we do not use PMDK libraries or any other extensive inner-code modifications, but rather employ a transparent approach in which scientific applications access NVM mainly using persistent memory file systems and conventional tools for HPC Checkpoint/Restart.

Our experimental environment consists of a dual-socket server (see specifications in Fig.~\ref{specs}). The DCPMM population configuration is 2-1-1 (see Fig.~\ref{population}). 
When using NVM as a node’s memory expansion, we applied the Optane™ DCPMM Memory mode, 
while in the rest of the paper, we applied the App Direct mode.

\subsection*{Optane™ DCPMM as a Persistent Storage - FIO Benchmark}
\noindent The Flexible I/O Tester (FIO) \cite{axboe2014fio} is a versatile storage benchmark tool that is used both for benchmarking and stress/hardware verification. Fig.~\ref{FIO} shows a comparison between the bandwidth of basic read/write file operations in FIO on Optane™ DCPMM and SATA-SSD. We run FIO v3.7 using the \textit{sync} ioengine, with a 512MB file size per thread and 8KB read or write block size. For write workloads, we issue an \textit{fsync()} after each 8KB are written. Each workload runs for 30 seconds, and the number of threads vary from one to eight, with each thread accessing a different file. Like Izraelevitz et al.~\cite[Chapter~5.2]{izraelevitz2019basic}, we evaluate the performances using various file systems. We note that file systems which enable DAX cannot run on SATA-SSD, hence only the non-DAX file systems were tested on the SATA-SSD device. The trends of our results differ somewhat from Izraelevitz's results, apparently due to differences in system configuration and memory population. Nevertheless, similarly to~\cite{izraelevitz2019basic}, the results demonstrate that Optane™ DCPMM is superior to SSD with regards to \textit{basic I/O operations}. The rest of this work compares the performance of \textit{scientific HPC benchmarks}, described next, on Optane™ DCPMM and on SSD. 

\section{Scientific Computing Benchmarks}
\label{Chapter:Benchmarks}


\begin{figure*}[ht!]
\centering
\begin{subfigure}{0.01\textwidth}
\begin{turn}{90} 
GOps$/$sec (per process)
\end{turn} 
\end{subfigure}%
~
\begin{subfigure}[t!]{0.495\textwidth}
\centering
\scalebox{.9}{
\begin{tikzpicture}
\begin{axis}[
    xlabel={Memory Usage (GB)},
    xmin=0.05, xmax=1500,
    ymin=0, ymax=3,
    xtick={0.125, 0.25, 0.5, 1,2,4,8,16,32,64,128,256,512,1024},
    xticklabels= {\footnotesize{$2^{-3}$},\footnotesize{$2^{-2}$}, \footnotesize{$2^{-1}$},\footnotesize{$2^0$},\footnotesize{$2^1$},\footnotesize{$2^2$},\footnotesize{$2^3$},\footnotesize{$2^4$},\footnotesize{$2^5$},\footnotesize{$2^6$},\footnotesize{$2^7$},\footnotesize{$2^8$},\footnotesize{$2^9$},\footnotesize{$2^{10}$}},
    xmode=log,
    legend pos=north east,
    ymajorgrids=true,
    grid style=dashed,
    width=10cm,
    height=7cm,
    legend style={nodes={scale=0.75, transform shape}},
    extra x ticks={192},
    extra x tick labels={},
    extra tick style={grid=major,},
]
\addplot[
    color=blue,
    mark=*,
    mark size=2.5pt,
    ]
    coordinates {
    (0.2,1.58)(0.5,1.58)(1.8,1.64)(25.7,1.66)(383.7,0.721)
    };
\addplot[
    color=red,
    mark=square*,
    ]
    coordinates {
    (0.1,1.55)(0.26,1.57)(0.83,1.65)(11.1,1.65)(166.6,1.28)
    };
\addplot[
    color=green,
    mark=pentagon*,
    mark size=3pt,
    ]
    coordinates {
    (0.162,0.87)(0.454,0.87)(1.476,0.87)(21.2,0.82)(318.9,0.374)
    };
\addplot[
    color=purple,
    mark=diamond*,
    mark size=3pt,
    ]
    coordinates {
    (0.06,2.4)(0.5,2.58)(3.58,2.21)(27.68,2.01)(218.4,0.391)
    };
\addplot[
    color=orange,
    mark=triangle*,
    mark size=3pt,
    ]
    coordinates {
    (0.46,1.55)(1.85,1.66)(7.23,1.57)(115,1.39)(919.5,0.11)
    };
    \legend{BT,LU,SP,MG,FT}
\end{axis}
\end{tikzpicture}
}
\caption{DCPMM Expansion (Optane™ DCPMM in Memory Mode)}
\end{subfigure}%
~ 
\begin{subfigure}[t!]{0.495\textwidth}
\centering
\scalebox{.9}{
\begin{tikzpicture}
\begin{axis}[
    xlabel={Memory Usage (GB)},
    xmin=0.05, xmax=1500,
    ymin=0, ymax=3,
    xtick={0.125, 0.25, 0.5, 1,2,4,8,16,32,64,128,256,512,1024},
    xticklabels= {\footnotesize{$2^{-3}$},\footnotesize{$2^{-2}$}, \footnotesize{$2^{-1}$},\footnotesize{$2^0$},\footnotesize{$2^1$},\footnotesize{$2^2$},\footnotesize{$2^3$},\footnotesize{$2^4$},\footnotesize{$2^5$},\footnotesize{$2^6$},\footnotesize{$2^7$},\footnotesize{$2^8$},\footnotesize{$2^9$},\footnotesize{$2^{10}$}},
    xmode=log,
    legend pos=north east,
    ymajorgrids=true,
    grid style=dashed,
    width=10cm,
    height=7cm,
    legend style={nodes={scale=0.75, transform shape}},
    extra x ticks={192},
    extra x tick labels={},
    extra tick style={grid=major,},
]
    
\addplot[
    color=blue,
    mark=*,
    mark size=2.5pt,
    ]
    coordinates {
    (0.2,1.58)(0.5,1.604)(1.8,1.649)(25.7,1.675)(383.7,0.0091)
    };
\addplot[
    color=red,
    mark=square*,
    ]
    coordinates {
    (0.1,1.505)(0.26,1.596)(0.83,1.673)(11.1,1.718)(166.6,1.697)
    };
\addplot[
    color=green,
    mark=pentagon*,
    mark size=3pt,
    ]
    coordinates {
    (0.162,0.887)(0.454,0.873)(1.476,0.879)(21.2,0.830)(318.9,0.0083)
    };
\addplot[
    color=purple,
    mark=diamond*,
    mark size=3pt,
    ]
    coordinates {
    (0.06,2.339)(0.5,2.531)(3.58,2.411)(27.68,2.248)(218.4,0.025)
    };
\addplot[
    color=orange,
    mark=triangle*,
    mark size=3pt,
    ]
    coordinates {
    (0.46,1.564)(1.85,1.663)(7.23,1.586)(115,1.404)
    };
    \legend{BT,LU,SP,MG,FT}
\end{axis}
\end{tikzpicture}
}
\caption{Memory Swap (SATA-SSD swap partition)}
    \end{subfigure}
    \caption{Floating-point Operation Throughput of NPB Benchmarks}
    \label{Fig:NPB}
\end{figure*}
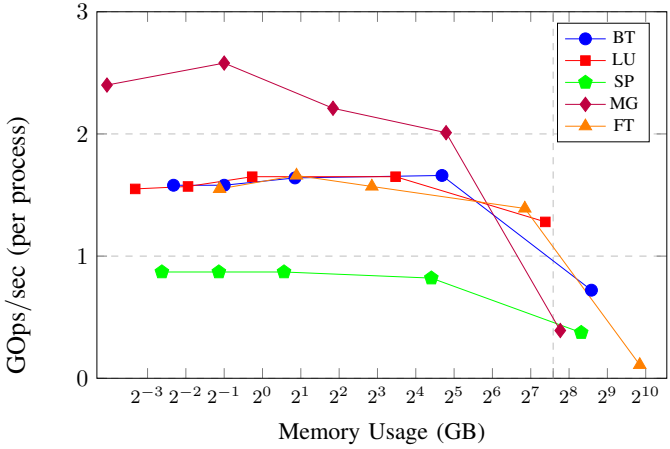
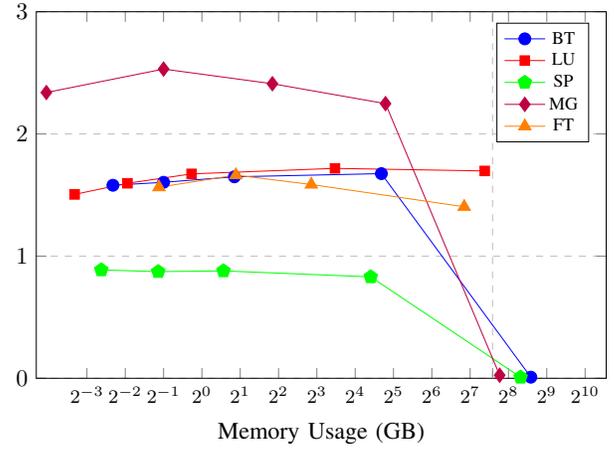
\begin{figure*}[ht!]
\centering
\begin{subfigure}{0.01\textwidth}
\begin{turn}{90} 
GOps$/$sec
\end{turn} 
\end{subfigure}%
~
\begin{subfigure}[t!]{0.495\textwidth}
\centering
\scalebox{.9}{
\begin{tikzpicture}
\begin{axis}[
    xlabel={Memory Usage (GB)},
    xmin=0, xmax=450,
    ymin=0, ymax=3.3,
    xtick= {0,50,100,150,200,250,300,350,400,450},
    ytick={0.5,1,1.5,2,2.5,3},
    legend pos=north east,
    ymajorgrids=true,
    grid style=dashed,
    width=10cm,
    height=7cm,
    legend style={nodes={scale=0.75, transform shape}},
    extra x ticks={96},
    extra x tick labels={},
    extra tick style={grid=major,},
]

\addplot[
    color=blue,
    mark=*,
    mark size=2.5pt,
    ]
    coordinates {
    (2,2.327)(4.4,2.327)(7.8,2.301)(12.3,2.319)(17.6,2.080)(24,2.256)(31.3,2.182)(39.64,1.990)(48.94,2.097)(59.2,2.047)(70.4,1.880)(82.7,1.943)(95.9,2.005)(110.1,1.598)(125.2,1.335)(141.4,1.168)(158.5,1.076)(176.6,0.994)(195.8,0.949)(215.8,1.012)(236.9,1.012)(258.9,1.007)(282.5,0.981)(305.9,0.980)(330.9,0.940)(356.8,0.941)(383.7,0.948)(411.6,0.950)(440.5,0.948)
    };

\addplot[
    color=purple,
    mark=diamond*,
    mark size=3pt,
    ]
    coordinates {
   (1,1.996)(3.9,2.019)(8.8,1.743)(15.6,1.703)(24.5,1.668)(35.2,1.694)(48,1.707)(62.6,1.616)(79.3,1.698)(97.9,1.691)(118.4,0.904)(140.9,0.663)(165.4,0.541)(191.8,0.489)(220.2,0.477)(250.9,0.468)(282.8,0.475)(317,0.472)(353.2,0.480)(391.4,0.475)(431.5,0.466)
    };
    \addplot[
    color=cyan,
    mark=x,
    mark size=3pt,
    ]
    coordinates {
  (1.5,2.375)(5.9,2.409)(13.2,2.334)(23.5,2.322)(36.7,2.178)(52.9,2.154)(72,2.174)(93.9,2.153)(118.9,1.173)(146.8,0.849)(177.6,0.704)(211.4,0.63)(248.1,0.600)(287.7,0.595)(330.3,0.606)(375.8,0.609)(424.2,0.603)
    };
      \addplot[
    color=orange,
    mark=triangle*,
    mark size=3pt,
    ]
    coordinates {
  (2,2.287)(3.868,2.301)(6.6,2.291)(10.547,2.232)(15.7,2.116)(22.385,2.277)(30.6,2.285)(40.849,2.250)(53,2.206)(67.373,2.116)(84.1,2.113)(103.455,2.010)(125.5,1.825)(150.584,1.704)(178.7,1.643)(210.165,1.597)(245.1,1.578)(283.705,1.599)(326.1,1.606)(372.639,1.605)(423.4,1.587)
    };
    \addplot[
    color=green,
    mark=pentagon*,
    mark size=3pt,
    ]
    coordinates {
 (1.5,1.252)(5.9,1.235)(13.2,1.215)(23.5,1.192)(36.7,1.180)(52.9,1.184)(71.9,1.184)(93.9,1.173)(118.9,0.896)(146.8,0.730)(178,0.640)(211.8, 0.609)(248.5,0.598)(288.2,0.593)(331,0.589)(352,0.591)(374,0.592)(398,0.588)
    };
    \addplot[
    color=red,
    mark=square*,
    ]
    coordinates {
 (2,0.696)(4.4,0.611)(12.2,0.576)(24,0.570)(39.6,0.526)(48.9,0.536)(59.2,0.529)(70.4,0.514)(82.7,0.5)(95.9,0.507)(110.1,0.442)(125.2,0.258)(141.4,0.365)(158.5,0.344)(176.6,0.329)(195.7,0.283)(215.8,0.312)(236.8,0.300)(258.8,0.304)(282.4,0.240)(305.8,0.288)(330.7,0.285)(352,0.285)(376,0.272)(399,0.289)
    };
        \addplot[
    color=black,
    mark=star,
    mark size=3pt
    ]
    coordinates {
   (0.4,1.727)(0.5,1.465)(3.8,1.465)(9,1.48)(15.6,1.43)(24.7,1.49)(34.2,1.499)(48,1.5)(63.2,1.41)(79,1.46)(97.8,1.42)(116,0.888)(141.9,0.56)(164.7,0.45)(192.9,0.44)(220.4,0.44)(250.7,0.425)(283.5,0.426)(315,0.423)(353,0.43)(391.6,0.446)
    };
    \legend{atax,jacobi-2d-imper,fdtd-2d, fdtd-apml, adi, gemver, covariance}
\end{axis}
\end{tikzpicture}
}
\caption{DCPMM Expansion (Optane™ DCPMM in Memory Mode)}
\end{subfigure}%
    ~ 
\begin{subfigure}[t!]{0.495\textwidth}
\centering
\scalebox{.9}{
\begin{tikzpicture}
\begin{axis}[
    xlabel={Memory Usage (GB)},
    xmin=0, xmax=250,
    ymin=0, ymax=3.3,
    xtick= {0,50,100,150,200,250},
    ytick={0.5,1,1.5,2,2.5,3},
    legend pos=north east,
    ymajorgrids=true,
    grid style=dashed,
    width=10cm,
    height=7cm,
    legend style={nodes={scale=0.75, transform shape}},
    extra x ticks={96},
    extra x tick labels={},
    extra tick style={grid=major,},
]

\addplot[
    color=blue,
    mark=*,
    mark size=2.5pt,
    ]
    coordinates {
    (2,2.01)(4.4,2.04)(7.8,1.98)(12.3,2.01)(17.6,1.98)(24,1.91)(31.3,1.86)(39.64,1.92)(48.94,1.90)(59.2,1.88)(70.4,1.85)(82.7,1.86)(95.9,0.05)(110.1,0.05)(125.2,0.05)(141.4,0.048)(158.5,0.049)(176.6,0.046)(195.8,0.045)(215.8,0.0433)(236.9,0.0458)
    };

\addplot[
    color=purple,
    mark=diamond*,
    mark size=3pt,
    ]
    coordinates {
   (1,1.71)(3.9,1.60)(8.8,1.51)(15.6,1.57)(24.5,1.52)(35.2,1.53)(48,1.53)(62.6,1.50)(79.3,1.54)(97.9,0.01)(118.4,0.01)(140.9,0.012)(165.4,0.009)(191.8,0.01)
    };
    \addplot[
    color=cyan,
    mark=x,
    mark size=3pt,
    ]
    coordinates {
  (1.5,2.447)(5.9,2.424)(13.2,2.433)(23.5,2.411)(36.7,2.297)(52.9,2.266)(72,2.258)(93.9,0.035)(118.9,0.037)(146.8,0.037)(177.6,0.036)
    };
      \addplot[
    color=orange,
    mark=triangle*,
    mark size=3pt,
    ]
    coordinates {
  (2,2.287)(3.868,2.301)(10.547,2.294)(22.385,2.295)(40.849,2.297)(67.373,2.296)(103.455,1.540)(150.584,0.111)(210.165,0.117)
    };
    \addplot[
    color=green,
    mark=pentagon*,
    mark size=3pt,
    ]
    coordinates {
 (1.5,1.22)(5.9,1.21)(13.2,1.19)(23.5,1.19)(36.7,1.17)(52.9,1.17)(71.9,1.17)(93.9,0.008)(118.9,0.0089)(146.8,0.0086)(178,0.0079)(211.8,0.008)
    };
    \addplot[
    color=red,
    mark=square*,
    ]
    coordinates {
 (2,0.609)(4.4,0.544)(12.2,0.572)(24,0.554)(39.6,0.536)(48.9,0.532)(59.2,0.522)(70.4,0.503)(82.7,0.494)(95.9,0.047)(110.1,0.037)(125.2,0.031)(141.4,0.022)(158.5,0.018)
    };
    \addplot[
    color=black,
    mark=star,
    mark size=3pt
    ]
    coordinates {
  (0.5,1.51)(3.8,1.465)(9,1.484)(15.6,1.46)(24.7,1.488)(34.2,1.499)(48,1.47)(63.2,1.49)(79,1.46)(97.8,0.01)(116,0.011)(141.9,0.009)(164.7,0.01)
    };
    \legend{atax,jacobi-2d-imper,fdtd-2d, fdtd-apml, adi, gemver, covariance}
\end{axis}
\end{tikzpicture}
}
\caption{Memory Swap (SATA-SSD swap partition)}
\end{subfigure}
    \caption{Floating-point Operation Throughput of PolyBench Benchmarks}
    \vspace{-0.5cm}
    \label{Fig:polybench}
\end{figure*}
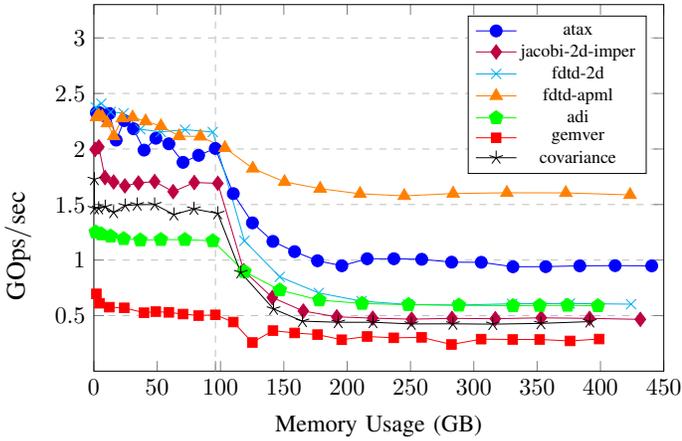
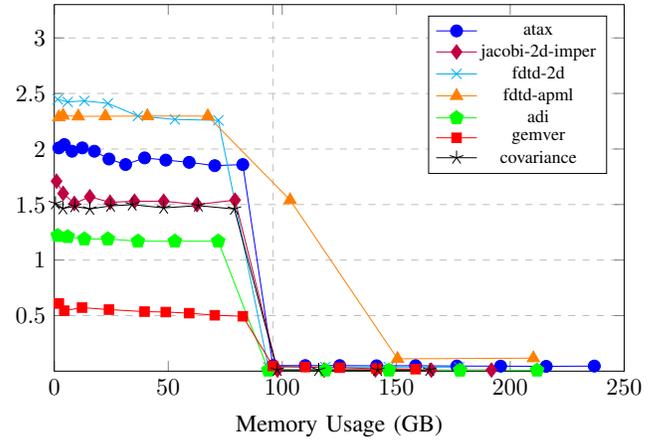

The NPB~\cite{bailey1991parallel} and PolyBench 4.2.1 beta~\cite{polybench} 
parallel benchmarks are used in this work in order to evaluate how scientific applications perform on storage-free setting and to establish a comparable performance metric.
PolyBench was adjusted to the present memory hierarchy,
so that the DRAM acts as a cache to the DCPMM. 
All of our benchmarks were executed using a single computation node 
(see the specs in Section~\ref{Chapter:NVRAM}). 

The Numerical Aerodynamics Simulations (NAS) Parallel Benchmarks set is a well-known suite of applications, developed by NASA, designed to evaluate the performance of high-performance computers. Originally, NAS Parallel Benchmarks (NPB) consisted of five parallel kernels: Integer Sort (IS), Embarrassingly Parallel (EP), Conjugate Gradient (CG), Multi-Grid (MG) and Fourier Transform (FT); and three pseudo-applications: Block Tri-diagonal solver (BT), Scalar Penta-diagonal solver (SP) and Lower-Upper Gauss-Seidel solver (LU)~\cite{bailey2011parallel}. 
Later, a few other representative HPC applications were added to the benchmark suite,
e.g., Unstructured Adaptive mesh (UA), BT using parallel I/O techniques (BTIO). 
Problem sizes in NPB are predefined and classified as follows: 
small size (class S), 90's workstation size (class W), 
standard sizes with ~$\times4$ size increase between the following classes (classes A,B,C), and large sizes with~$\times16$ size increase (classes D,E,F).

A prominent benchmark is BT, simulating a computational fluid dynamics application.
Its solution is derived by solving three uncoupled systems of equations: 
first at the direction of $x$, then at the direction of $y$, 
and finally at the direction of $z$.
These systems are composed of $5\times5$ block matrices given in a block-tridiagonal form. 
We chose to concentrate on this benchmark throughout our work. 
We also used a derivative of BT, BTIO, 
which benchmarks the performance of PnetCDF and MPI-IO methods for 
the I/O pattern used by the NPB suite. 
It measures the communication speeds for parallel intensive I/O, 
and is used for evaluating storage performance.




It is not easy to modify the predefined problem sizes in NPB.
As can be seen in Fig.\ref{Fig:NPB}, the differences between them are too coarse around the capacity of DRAM that we use in a node (192GB). 
To evaluate the memory expansion of Optane™ DCPMM, a more refined examination of the hardware behavior is needed around the point where DRAM is fully occupied. We use the PolyBench suite for this purpose, 
since its granularity can be tuned in finer manner than NPB.
PolyBench~\cite{polybench, yuki2014understanding} 
is a collection of 30 representative HPC applications from various domains,
e.g., linear algebra, image processing, physics simulation. 
PolyBench aims to make the execution of the kernels as uniform as possible, 
using benchmarks that contain static control parts. 
Kernel setting is done by a single file that is tuned at compile time. 
This file executes supplementary operations such as cache flushing 
before the kernels are executed, 
and can carry out real-time scheduling to prevent operating-system interference. 

PolyBench 4.2.1 has five predefined problem sizes: 
less than 16KB of memory---may fit within L1 (MINI), 
around 128KB of memory---may fit within L1 (SMALL), 
around 1MB of memory---would not fit in L1, but may fit within L2 (MEDIUM), 
around 25MB of memory---would not fit in L3 (LARGE),
and finally, around 120MB of memory (EXTRALARGE). 
Although PolyBench was designed to evaluate the performance of HPC 
applications on small predefined problem sizes, 
these problem sizes can be easily modified. 
In order to further test the performance of HPC applications coupled with non-volatile memory, we modified the benchmarks 
so that the behavior of the execution is examined relative to 
the memory consumption in a finer granularity, 
specifically around the DRAM capacity which is 96GB in a socket of our setting (we bind PolyBench to memory resources of one socket).
Most of the benchmarks in the same category are computationally comparable (e.g jacobi-1d versus jacobi-2d). Therefore, we chose representative benchmarks in each category (except for \textit{Medley}, considered redundant in this context), while focusing especially on benchmarks from \textit{Linear-Algebra} and \textit{Stencils}. 
When technically possible, for example in stencils computations, 
we executed the benchmarks for a relatively small number of time steps, 
not necessarily reaching convergence.

\begin{figure*}[ht!]
\begin{subfigure}{0.01\textwidth}
\begin{turn}{90} 
Total MiB$/$sec
\end{turn} 
\end{subfigure}%
~
\begin{subfigure}[t!]{0.495\textwidth}
\centering
\scalebox{.9}{
\begin{tikzpicture}
\begin{axis}[
    xlabel={Total I/O amount (GB)},
    xmin=0, xmax=100,
    ymin=0, ymax=800,
    xtick= {0.00125,0.01,0.08,0.64,5.12,40.96},
    xticklabels= {0.00125,0.01,0.08,0.64,5.12,40.96},
    ytick={0,200,400,600,800},
    legend pos=north west,
    ymajorgrids=true,
    grid style=dashed,
    width=10cm,
    height=7cm,
    xmode=log,
    legend style={nodes={scale=0.75, transform shape}}
]

\addplot[
    color=blue,
    mark=*,
    mark size=2.5pt,
    ]
    coordinates {
    (0.00125,22.46)(0.01,112.86)(0.08,539.97)(0.64,655.73)(5.12,726.58)(40.96,731.39)
    };

\addplot[
    color=purple,
    mark=diamond*,
    mark size=3pt,
    ]
    coordinates {
    (0.00125,7.79)(0.01,122.83)(0.08,364.2)(0.64,575.72)(5.12,625.65)(40.96,655.74)
    };
    \addplot[
    color=red,
    mark=square*,
    ]
    coordinates {
    (0.00125,19.84)(0.01,105.87)(0.08,298.42)(0.64,387.57)(5.12,406.45)(40.96,413.45)
    };
    \addplot[
    color=orange,
    mark=triangle*,
    mark size=3pt,
    ]
    coordinates {
    (0.00125,12.25)(0.01,86.76)(0.08,173.42)(0.64,206.69)(5.12,208.95)(40.96,181.16)
    };
    \addplot[
    color=green,
    mark=pentagon*,
    mark size=3pt,
    ]
    coordinates {
    };
    \legend{SplitFS (NVM),NOVA (NVM),ext4-dax (NVM), xfs (SATA SSD)}
\end{axis}
\end{tikzpicture}
}
\caption{BTIO MPI Collective I/O (Write)}
\end{subfigure}%
~
\begin{subfigure}[t!]{0.495\textwidth}
\centering
\scalebox{.9}{
\begin{tikzpicture}
\begin{axis}[
    xlabel={Total I/O amount (GB)},
    xmin=0, xmax=100,
    ymin=0, ymax=2500,
    xtick= {0.00125,0.01,0.08,0.64,5.12,40.96},
    xticklabels= {0.00125,0.01,0.08,0.64,5.12,40.96},
    ytick={0,500,1000,1500,2000,2500},
    legend pos=north west,
    ymajorgrids=true,
    grid style=dashed,
    width=10cm,
    height=7cm,
    xmode=log,
    legend style={nodes={scale=0.75, transform shape}}
]

\addplot[
    color=blue,
    mark=*,
    mark size=2.5pt,
    ]
    coordinates {
    (0.00125,23.62)(0.01,113.86)(0.08,542.96)(0.64,707.271)(5.12,710.51)(40.96,732.36)
    };

\addplot[
    color=purple,
    mark=diamond*,
    mark size=3pt,
    ]
    coordinates {
    (0.00125,10.07)(0.01,55.07)(0.08,186.06)(0.64,324.14)(5.12,460.81)(40.96,609.41)
    };
    \addplot[
    color=red,
    mark=square*,
    ]
    coordinates {
    (0.00125,19.78)(0.01,80.93)(0.08,180.9)(0.64,254.44)(5.12,296.55)(40.96,369.49)
    };
      \addplot[
    color=orange,
    mark=triangle*,
    mark size=3pt,
    ]
    coordinates {
    (0.00125,11.92)(0.01,72.96)(0.08,116.25)(0.64,129.33)(5.12,157.15)(40.96,226.2)
    };
    \addplot[
    color=green,
    mark=pentagon*,
    mark size=3pt,
    ]
    coordinates {
    };
    \legend{SplitFS (NVM),NOVA (NVM),ext4-dax (NVM), xfs (SATA SSD)}
\end{axis}
\end{tikzpicture}
}
\caption{BTIO MPI Independent I/O (Write)}
\end{subfigure}%
\vskip\baselineskip
\begin{subfigure}{0.01\textwidth}
\begin{turn}{90} 
Total MiB$/$sec
\end{turn} 
\end{subfigure}%
~
\begin{subfigure}[t!]{0.495\textwidth}
\centering
\scalebox{.9}{
\begin{tikzpicture}
\begin{axis}[
    xlabel={Total I/O amount (GB)},
    xmin=0, xmax=100,
    ymin=0, ymax=4000,
    xtick= {0.00125,0.01,0.08,0.64,5.12,40.96},
    xticklabels= {0.00125,0.01,0.08,0.64,5.12,40.96},
    ytick={0,1000,2000,3000,4000},
    legend pos=north west,
    ymajorgrids=true,
    grid style=dashed,
    width=10cm,
    height=7cm,
    xmode=log,
    legend style={nodes={scale=0.75, transform shape}}
]

\addplot[
    color=blue,
    mark=*,
    mark size=2.5pt,
    ]
    coordinates {
    (0.00125,25.68)(0.01,175.18)(0.08,956.13)(0.64,822.63)(5.12,1587.56)(40.96,1623.31)
    };

\addplot[
    color=purple,
    mark=diamond*,
    mark size=3pt,
    ]
    coordinates {
    (0.00125,22.84)(0.01,137.4)(0.08,411.98)(0.64,948.58)(5.12,1104.81)(40.96,1090.78)
    };
    \addplot[
    color=red,
    mark=square*,
    ]
    coordinates {
    (0.00125,22.26)(0.01,139.23)(0.08,503.66)(0.64,1011.36)(5.12,1189.24)(40.96,1223.54)
    };
      \addplot[
    color=orange,
    mark=triangle*,
    mark size=3pt,
    ]
    coordinates {
    (0.00125,19.08)(0.01,94.55)(0.08,240.28)(0.64,277.63)(5.12,298.23)(40.96,309.6)
    };
    \addplot[
    color=green,
    mark=pentagon*,
    mark size=3pt,
    ]
    coordinates {
    };
    \legend{SplitFS (NVM),NOVA (NVM),ext4-dax (NVM), xfs (SATA SSD)}
\end{axis}
\end{tikzpicture}
}
\caption{BTIO MPI Collective I/O (Read)}
\end{subfigure}%
~
\begin{subfigure}[t!]{0.495\textwidth}
\centering
\scalebox{.9}{
\begin{tikzpicture}
\begin{axis}[
    xlabel={Total I/O amount (GB)},
    xmin=0, xmax=100,
    ymin=0, ymax=4000,
    xtick= {0.00125,0.01,0.08,0.64,5.12,40.96},
    xticklabels= {0.00125,0.01,0.08,0.64,5.12,40.96},
    ytick={0,1000,2000,3000,4000},
    legend pos=north west,
    ymajorgrids=true,
    grid style=dashed,
    width=10cm,
    height=7cm,
    xmode=log,
    legend style={nodes={scale=0.75, transform shape}}
]

\addplot[
    color=blue,
    mark=*,
    mark size=2.5pt,
    ]
    coordinates {
    (0.00125,25.32)(0.01,178.89)(0.08,928.77)(0.64,1700.53)(5.12,1583.24)(40.96,1673.98)
    };

\addplot[
    color=purple,
    mark=diamond*,
    mark size=3pt,
    ]
    coordinates {
    (0.00125,15.67)(0.01,108.35)(0.08,606.36)(0.64,1659.01)(5.12,2376.99)(40.96,3698.66)
    };
    \addplot[
    color=red,
    mark=square*,
    ]
    coordinates {
    (0.00125,21.91)(0.01,182.32)(0.08,925.07)(0.64,1813.38)(5.12,2801.88)(40.96,3752.81)
    };
      \addplot[
    color=orange,
    mark=triangle*,
    mark size=3pt,
    ]
    coordinates {
    (0.00125,22.66)(0.01,115.61)(0.08,224.64)(0.64,293.53)(5.12,331.02)(40.96,330.32)
    };
    \addplot[
    color=green,
    mark=pentagon*,
    mark size=3pt,
    ]
    coordinates {
    };
    \legend{SplitFS (NVM),NOVA (NVM),ext4-dax (NVM), xfs (SATA SSD)}
\end{axis}
\end{tikzpicture}
}
\caption{BTIO MPI Independent I/O (Read)}
\end{subfigure}%
    \caption{Memory Bandwidth of BTIO Benchmark with the MPI-IO method}
    \vspace{-0.5cm}
    \label{btio}
\end{figure*}
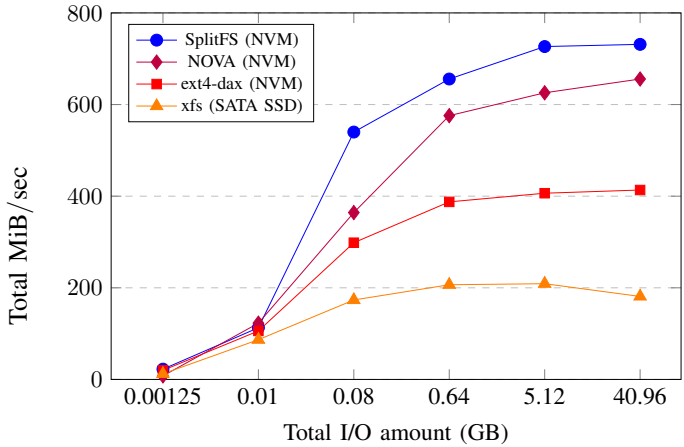
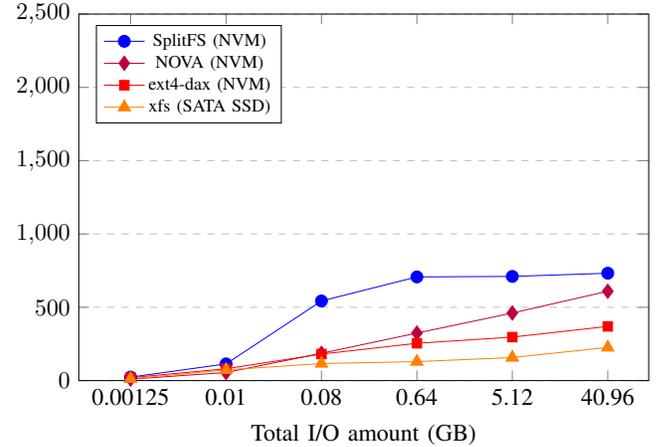
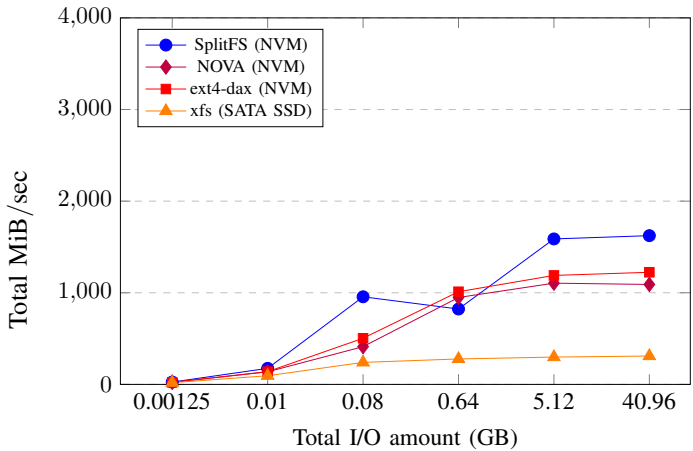
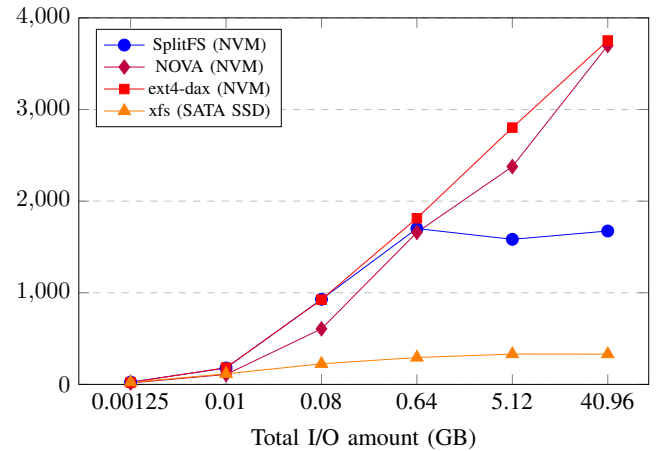




\section{DCPMM as Extended Virtual Memory}
\label{Chapter:ExtendedVirtualMemory}


A big limitation of traditional HPC is the main memory capacity,
i.e., the size of DRAM that is installed in a compute node is limited due to hardware, software and economic factors. 
In order to overcome DRAM limits some HPC applications distribute their memory (and computations) between several compute nodes (e.g. using MPI~\cite{gabriel2004open}), or use the virtual memory mechanism, which swaps the contents to or from the primary memory with the secondary storage. Both solutions have disadvantages: distributing memory among multiple nodes comes with programming efforts but mainly large overheads of communication between the nodes; and swapping data to a secondary memory dramatically decreases performances due to extremely high access time to the secondary storage (compared to DRAM) and to kernel management overheads. 
Using NVM as an extension to the primary memory creates a solution to the DRAM limit problem, which is transparent to the operating system and to the user. 

When configuring Optane™ DCPMM in Memory Mode as primary memory extension on a node, DRAM DIMMs are logically viewed as L4 cache and not as system memory. Therefore the total system memory as seen by the operating system is based on the total capacity of the installed DCPMM devices~\cite{tristian2019analyzing}. To evaluate the usage of DCPMM devices as primary memory expansion, NPB and PolyBench benchmarks were executed on a single node of our experimental environment (Section~\ref{Chapter:NVRAM}), with all its Optane™ DCPMMs configured in Memory Mode. PolyBench benchmarks were executed on one core with the memory 
resources available in one socket (96GB DRAM and 512GB Optane™ DCPMM). 
NPB benchmarks were executed with 16 processes using the entire memory 
resources of a node (192GB DRAM and 1024GB Optane™ DCPMM). 
To compare these results with a traditional solution, these benchmarks 
were also executed on a configuration without Optane™ DCPMMs, but with memory swapping to a SATA-SSD device. 
As can be seen in Figs.~\ref{Fig:NPB} and~\ref{Fig:polybench}, the performances of scientific applications are almost the same before passing the DRAM limit (marked in gray line). After passing the DRAM limit, however, they decrease only by a factor of 2-3 on average when using Optane™ DCPMMs for primary memory extension, whereas they deteriorate by 2-3 orders of magnitude when using memory swapping to SSD. This serves as evidence to the dramatic improvements in performance of scientific applications that Optane™ DCPMMs offers for large workloads. 

\section{DCPMM as Persistent-Memory for Local and Distributed File Systems}
\label{Chapter:DistributedStorage}

\begin{figure*}[ht]
\centering
\begin{subfigure}[t]{0.32\textwidth}
\centering
\scalebox{.8}{
\begin{tikzpicture}
\begin{axis}[
    xlabel={Number of processes},
    ylabel= {Time (seconds)},
    ylabel near ticks,
    xmin=0.9, xmax=16,
    xtick= {1,4,9,16,25},
    ymajorgrids=true,
    grid style=dashed,
    width=8cm,
    legend style={nodes={scale=0.9, transform shape}, at={(1,1)},anchor=north east}
]

    \addplot[
    color=blue,
    mark=*,
    mark size=2.5pt,
    ]
    coordinates {
      (1,19598.35)(4,4361.5)(9,2221.63)(16,1244.27)
    };

    \addplot[
    color=purple,
    mark=diamond*,
    mark size=3pt,
    ]
    coordinates {
      (1,29324.48)(4,9093.11)(9,4985.81)(16,4651.84)
    };
    \addplot[
    color=red,
    mark=square*,
    ]
    coordinates {
      (1,25475.93)(4,10647.99)(9,7853.09)(16,8061.17)
    };
    \addplot[
    color=green,
    mark=pentagon*,
    mark size=3pt,
    ]
    coordinates {
      (1,33111.65)(4,15553.69)(9,12988.97)(16,12223.27)
    };
    \addplot[
    color=orange,
    mark=triangle*,
    mark size=3pt,
    ]
    coordinates {
      (1,26062.18)(4,18006.65)(9,19216.75)(16,20073.69)
    };
    \legend{BT-NVM,BT-SCR-NVM,BT-DMTCP-NVM, BT-SCR-SSD, BT-DMTCP-SSD}
\end{axis}
\end{tikzpicture}
}
\caption{Crash-free BT on different non-volatile hardware}
\label{Fig:bt-crash-free}
\end{subfigure}%
    ~ 
\begin{subfigure}[t]{0.32\textwidth}
\centering
\scalebox{.8} {
\begin{tikzpicture}
\pgfplotsset{
every axis/.style={
    xlabel={Number of processes},
    ybar stacked,
    ymin=0, ymax=175,
    ymajorgrids=true,
    grid style=dashed,
    xtick= {1 ,4, 9, 16},
    ytick={25, 50, 75, 100, 125, 150, 175},
    bar width=5pt,
    width=8cm,
  },
}

\begin{axis}[bar shift=-10pt,hide axis]

\addplot coordinates {(1,4.85) (4,1.32) (9, 0.64) (16, 0.38) }; 
\addplot coordinates {(1,30.33802846) (4, 15.20276098) (9,9.744902335)  (16, 13.26159897) }; 
\addplot coordinates {(1,79.51009007) (4, 17.57899255) (9,8.946211879) (16, 4.984646252) }; 
\legend{Recovery, Checkpoint, Compute}

\addlegendentry{BT-SCR-NVM}
\addlegendimage{draw=black}

\addlegendentry{BT-SCR-SSD}
\addlegendimage{draw=black, pattern=north east lines}

\addlegendentry{BT-DMTC-NVRAM}
\addlegendimage{draw=black, pattern=horizontal lines}

\addlegendentry{BT-DMTC-SSD}
\addlegendimage{draw=black, pattern=dots}
\end{axis}

\begin{axis}[bar shift=-3pt, hide axis,]
\addplot+[
    postaction={
        pattern=north east lines
    }
] coordinates { (1,4.82) (4,1.3) (9, 0.63) (16, 0.4) };
\addplot+[
    postaction={
        pattern=north east lines
    }
] coordinates {(1,49.81768141) (4, 41.75222764) (9,42.23876082)  (16,46.83227138) };
\addplot+[
    postaction={
        pattern=north east lines}
] coordinates {(1,78.78065928) (4, 17.71230607) (9, 9.015615608) (16, 5.034552257) }; 
\end{axis}

\begin{axis}[bar shift=4pt, hide axis]
\addplot+[
    postaction={
        pattern=horizontal lines}
] coordinates { (1, 19.611394) (4, 6.4337233) (9, 6.032835) (16, 8.812077) };
\addplot+[
    postaction={
        pattern=horizontal lines
    }
] coordinates {(1, 61.97231674) (4, 23.65924869) (9, 22.06882029)  (16, 29.15825543) };
\addplot+[
    postaction={
        pattern=horizontal lines
    }
] coordinates {(1, 78.06483846) (4, 17.92693322) (9, 9.438928878) (16, 5.189859331) };
\end{axis}

\begin{axis}[bar shift=11pt]
\addplot+[
    postaction={
        pattern=dots
    }
] coordinates { (1, 11.483868) (4, 3.994187) (9, 3.132661) (16, 4.220131) };  
\addplot+[
    postaction={
        pattern=dots
    }
] coordinates {(1, 22.6886898) (4, 84.06118703) (9, 96.4790623)  (16, 90.41409413) };
\addplot+[
    postaction={
        pattern=dots
    }
] coordinates {(1, 84.0787835) (4, 18.21913594) (9, 9.929496533) (16, 5.232131243) }; 
\end{axis}
\end{tikzpicture}
}
\caption{Breakdown of recovery, checkpoint and compute time per iteration in a single crash and recovery run of BT.}
\label{Fig:bt-ckpt-per-iteration}
\end{subfigure}
    ~ 
\begin{subfigure}[t]{0.32\textwidth}
\centering
\scalebox{.8}{
\begin{tikzpicture}
\begin{axis}[
    xlabel={Number of processes},
    ylabel= {Size (gigabytes)},
    ylabel near ticks,
    xmin=0.9, xmax=16,
    ymin=0, ymax=30,
    xtick= {1,2,4,9,16},
    ytick={5,10,15,20,25},
    legend pos=south east,
    ymajorgrids=true,
    grid style=dashed,
    width=8cm,
    legend style={nodes={scale=0.9, transform shape}}
]
\addplot[
    color=blue,
    mark=*,
    mark size=2.5pt,
    ]
    coordinates {
       (1,8.7751)(4,8.9509)(8.8,9.1298)(16,9.3119)
    };
\addplot[
    color=purple,
    mark=diamond*,
    mark size=3pt,
    ]
    coordinates {
       (1,23.6484)(4,24.341)(9,25.3348)(16,26.9708)
    };
 
    \legend{BT-SCR,BT-DMTCP}
\end{axis}
\end{tikzpicture}
}
\caption{Checkpoint size of BT with SCR and DMTCP}
\label{Fig:bt-ckpt-size}
\end{subfigure}

\caption{Comparison of SCR and DMTCP on the BT benchmark on different non-volatile hardware settings.}
\vspace{-0.5cm}
    
 
\end{figure*}
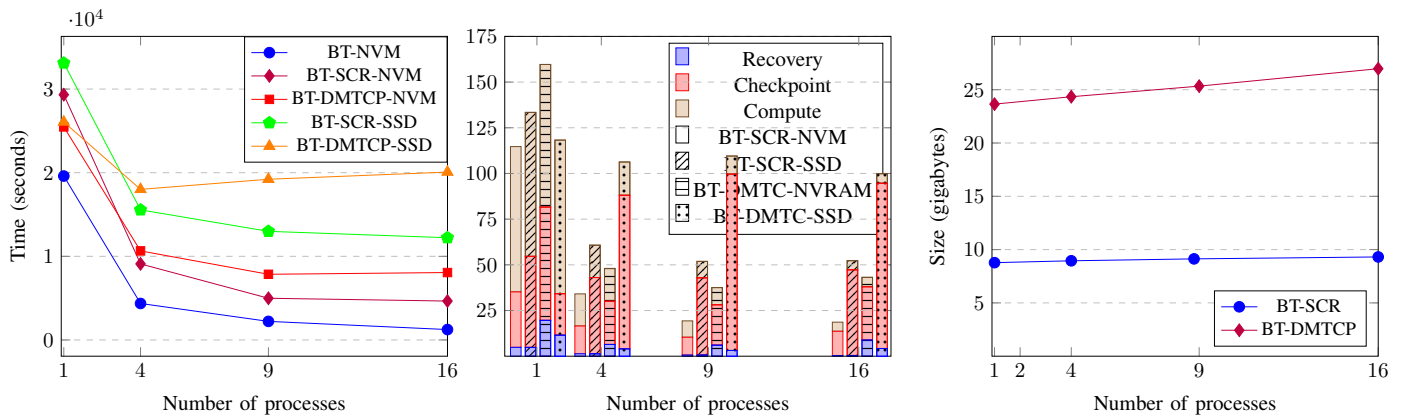

Persistent Memory File Systems (PMFSs) include both local and distributed file systems. 
Local persistent file systems include ext4, xfs (for Linux-based systems), 
and NTFS (for Windows-based systems). 
They support a special mode called Direct Access (DAX) that enables memory mapping directly from the NVM to the application memory space.
This bypasses the kernel, page cache, I/O subsystem, 
avoids interrupts and context switching, and allows the application to perform byte-addressable load/store memory operations~\cite{intel-quick-start-guide-persistent-memory}. 


Several other file systems have been developed with performance and other guarantees (e.g. consistency, atomicity, fault tolerance) in mind. 
These include NOVA~\cite{nova} and SplitFS~\cite{KadekodiEtAl19-SplitFS}, 
both providing POSIX compliant interfaces that support legacy scientific applications. 
NOVA is designed to maximize performance on hybrid memory (combining DRAM and NVM) systems while providing strong consistency guarantees. 
It is log-structured, and as such, it exploits the fast random access that NVMs provide, by storing the logs in the NVM and the indexes in the DRAM to allow fast search operations.
\emph{SplitFS} presents a split of responsibilities between a user-space library file system and an existing kernel PMFS. 
The user-space library file system handles data operations via POSIX calls interception, memory-mapping the underlying files, and serving the read and overwrites using processors loads and stores. 
The PM file system (ext4-dax) handles metadata operations.

While these local PMFSs are good choices for single-node workloads, scientific computing applications usually require a distributed PMFS that operates on several nodes.
Distributed PMFSs control how data is stored and retrieved from NVM when accessed by more than one node, using communication, such as Ethernet and RDMA.

Nowadays, most supercomputers contain disk-less nodes with a centralized storage, that all applications access and communicate with. This makes the formation of a distributed PMFS using distributed NVM units a great challenge. Octopus~\cite{octopus}, Assise~\cite{assise} and DAOS~\cite{daos-springer, daos-website} are the most contemporary and promising DFSs for this setting.

\emph{Octopus} closely couples NVM and RDMA in order to 
reduce memory copying overhead. 
It presents a directly-accessed shared persistent memory pool setup in every server, with the shared memory mechanism of Linux. This allows data fetching from clients, and data pushing to clients based on server and network loads. 
Octopus
is incompatible with Intel's Optane™ DCPMM since it manages its storage space on an emulated persistent memory of Linux in each server.

\emph{Assise} is built on the local file system Strata~\cite{strata}, originally developed as a cross-media file system.
Its main goal is to support work over several types of storage media (RAM, SSD, Disk), leveraging the strengths of one storage 
media to compensate for weaknesses of another. 
Assise extended this work to provide a DFS based on a persistent, replicated coherence protocol.
This protocol manages client-local Persistent-Memory as a linearizable and crash-recoverable cache between applications and slower 
(and possible remote) storage. 
When trying to use Assise, we discovered several issues.
First, its system call interception mechanism lacks support
for many necessary system calls. 
As a result, it is not possible to run even simple GNU programs such as \textit{touch} or \textit{rm}~\cite{assise-github-issue-7}. 
Second, Assise is incompatible with MPI implementations such as OpenMPI since the \textit{execve} system call 
is not yet supported~\cite{assise-github-issue-3-comment}.
Finally, DMTCP, which is widely used for checkpointing scientific applications cannot run with Assise since the system call interception mechanism of Assise collides with that of DMTCP's, resulting in an infinite loop.
Unlike Octopus and Assise, state-of-the-art \emph{DAOS} 
is designed from the ground up to support Storage Class 
Memory and NVMe storage in user space. 
Its advanced storage API enables the native support of structured, semi-structured, and unstructured data models, overcoming the limitations of traditional POSIX based parallel file-systems~\cite{daos-springer}. 
At the same time, DAOS provides POSIX access for legacy applications, as well as direct MPI-IO and HDF5 support. 
POSIX access is enabled by the \textit{libdfs} library 
that implements the file and directory abstractions 
over the native \textit{libdaos} library. 
In addition, there is a FUSE daemon (optionally
with an interception library, \textit{libioil}), 
which addresses some of the FUSE performance bottlenecks.
This delivers full OS bypass for POSIX read/write operations, and exposes the POSIX emulation transparently,
without any modifications~\cite{daos-springer, daos-website-posix}.

Fig.~\ref{btio} focuses on the BTIO benchmark with MPI-IO method to evaluate the I/O performances of some PMFSs on the Optane™ DCPMM in App Direct mode, in comparison to I/O performed on a SATA-SSD device. 
BTIO was executed with 4 processes, 
and was modified to execute a file sync (with \textit{MPI\_File\_sync}) 
after every write, to ensure the immediate transition of data to the Optane™ DCPMM and SATA-SSD devices. 
As can be seen, using Optane™ DCPMM with various PMFSs can dramatically increase I/O operations in scientific computing applications. The increase seems to be more significant on read workloads (up to $\times10$ speedup), compared to the increase of write workloads performances (up to $\times3$-$\times3.5$).

\section{DCPMM as Checkpoint/Restart (C/R) Storage}
\label{Chapter:CheckpointRestartStorage}

Many large scale, and especially exa-scale scientific simulations,
where performances is top priority, run on HPC platforms with millions of components.
These components are extremely vulnerable to failures, which leads to
the loss of simulation results 
as well as a waste of highly-priced resources 
(including power and actual computation time for other applications). 
This raises the need for a scalable and reliable fault tolerance 
mechanism that supports large scale scientific codes. 

In \emph{explicit C/R}, in order to successfully checkpoint and restart 
a program from a valid state, the program has to store 
(and restore from) the whole required computational state.
For scientific simulations, this usually consist of the data 
and the current time-step. 
The state is saved and restored manually within the program, and it 
is the responsibility of the programmer to understand what information 
is necessary and sufficient for correct recovery. 
If the state is represented compactly, i.e., the checkpoint file contains 
exactly what is required for a correct restart to be carried out, 
the C/R overhead might stay comparably low. 
\textit{Scalable Check-point/Restart (SCR)} is state-of-the-art
library~\cite{moody2010design,scrGit,scrDocs},
developed by Lawrence Livermore National Laboratory (LLNL),
which supports scalable explicit C/R. 
SCR deploys multilevel check-pointing~\cite{moody2010design}, 
which supports both 
(1) frequent, fast, but more volatile check-points to the node 
local RAM/Disk that may be used to recover from software faults, 
as well as (2) less frequent, slower but persistent check-points 
to the DFS that can withstand substantial software failures. 


In \emph{transparent C/R} the state of the program is saved without 
any knowledge of the application or intervention from the programmer. 
A possible implementation of such approach might be based on checkpointing 
in the userspace by catching and wrapping any needed system calls, 
in order to continually track the state of the program application. 
Additionally, since transparent C/R is oblivious to program behavior, 
each call that involves any external resources outside the process 
(e.g., network or storage), 
has to be documented via non-volatile log file in order to replay 
them in case of a failure. 
Capturing system calls and maintaining a log file inflicts a measurable 
performance penalty for the application.
The \emph{Distributed Multi-threaded CheckPointing} (DMTCP)~\cite{ansel2009dmtcp} 
library enables transparent C/R of a single-host, 
parallel or distributed computation, 
using a preloaded shared library that wraps system calls. 
DMTCP does so without any requirement for modifications to the source code 
or the operating system, supporting a variety of HPC languages and infrastructures, 
including MPI and OpenMP~\cite{rodriguez2019job}, and InfiniBand~\cite{cao_transparent_2013}. 

Restoring the state of the program requires writing to and reading from 
non-volatile hardware. 
Traditionally, programs and in particular HPC simulations carried 
out their C/R functionality in conventional storage (HDD and SSDs).
This imposed a significant running-time penalty. 
Since NVM offers a non-volatile 'storage' option with improved 
read and write stats, it is interesting to evaluate the performance 
gap between it and traditional storage.
In order to test the C/R performance on the different non-volatile hardware, 
we use the BT benchmark from the NAS Parallel Benchmarks suite~\cite{bailey1991parallel}. 
Note that the scientific application evaluated 
in Section~\ref{Chapter:ExtendedVirtualMemory} involves only writes to the NVM, 
while the following application consists of both writes to and reads from the NVM.

Fig.~\ref{Fig:bt-crash-free}-\ref{Fig:bt-ckpt-size} evaluate the cost of storing checkpoints by presenting the 
performance of three algorithms: 
Original BT (without check-pointing); BT with SCR; and BT with DMTCP, 
on the problem class D (matrix of size 408 $\times$ 408 $\times$ 408), 
on two non-volatile hardware components: SSD and DCPMM (as described in Section~\ref{Chapter:NVRAM}). 
Problem class D was chosen as it is the largest problem that fits into memory.

Similarly to Maroñas et al.~\cite{maronas2020extending}, 
BT's MPI implementation was expanded to support scalable checkpoint recovery (SCR).
That is, at the end of each iteration, the solver inner state is checkpointed. 
Upon recovery after a failure, the most recent valid checkpoint is loaded.
Due to the layered nature of SCR, our implementation flushes the checkpoint 
from the cache to the persistent memory, at the end of each iteration.
While DMTCP checkpoints are external to the process, for a fair comparison, 
we modified BT to initiate checkpoints at the end of each iteration.

Fig.~\ref{Fig:bt-ckpt-per-iteration} evaluates the cost of writes and reads to and from NVM and SSD in a scientific application. It measures the crash and recovery performance of BT with SCR and BT with DMTCP. Values in Fig.~\ref{Fig:bt-ckpt-per-iteration} are normalized to a single iteration.

Fig.~\ref{Fig:bt-crash-free} shows that in both scenarios, as the number of processes is close to 1, DMTCP implementations perform better than their SCR equivalents. A 15\% rise is seen in NVM and a 27\% increase in SSD in the total run time when running crash-free with a single process. We can see that the trend changes beyond 4 processes: as the number of processes increases, SCR performs better. SCR shows a 42.2\% decrease in run time on NVM and a decrease of 39.1\% on SSD when running with 16 processes. This can be explained by higher overhead times that SCR suffers from, when running as a single process. DMTCP scaled worse beyond 4 processes, leading to a worse processing time as the number of processes grows. Overall, both algorithms perform better on NVRAM than on SSD as it performs x1.02-x2.49, 1.12-2.62x faster on BT-DMTCP and on BT-SCR respectively.

In the settings of a single crash and recovery, BT-SCR shows a smaller recovery overhead than the recovery time of BT-DMTCP regardless of the number of processes. As the number of processes grows, the recovery overhead drops. BT-SCR scales better as the recovery time decreases. Specifically, the recovery in NVM and SSD seems equal on BT-SCR as the read operation is very short and therefore the difference seems insignificant. BT-DMTCP shows an improvement in up to 9 processes, where beyond 16 processes there is a slight increase in recovery duration of $\times 1.36$ and $\times 1.1$ for NVRAM and SSD respectively. This increase can be explained by the fact that the computation spans on two sockets, therefore it requires slower communication. Surprisingly, in the case of BT-DMTCP, the recovery performance on NVRAM is slightly worse than the performance on SSD (between $\times 1.5$-$\times 2$). 

The checkpoint duration demonstrates several trends. In most scenarios, both algorithms perform faster checkpointing on NVRAM than on SSD. BT-SCR shows shorter checkpoint durations over NVRAM, compared to SSD. As the number of processes increases to 9 threads, the checkpoint duration speeds up by $\times 3.11$, whereas with 16 processes it slows down by $\times 2.28$. However, on top of SSD, there is a mild speed-up from a serial run to 4 processes ($\times 1.19$) with a slight decrease beyond to $\times 1.06$. BT-DMTCP on top of NVRAM, shows a similar pattern as BT-SCR on top of NVRAM, with a checkpoint speed-up of $\times 2.8$ to 9 processes with a slow-down by $\times 2.12$ with 16 processes. On SSD however, we see a slow-down of $\times 0.25$ of the checkpoint duration from 1 process to 4 processes where beyond 4 processes there isn't any further slow-down. 
As expected, the computation time decreases as the number of processes grows, and remains oblivious to the hardware on which the checkpoint is written. 

Fig.~\ref{Fig:bt-ckpt-size} demonstrates the differences between conducting explicit C/R and implicit C/R in terms of checkpoint size in a scientific application, by measuring the checkpoint size of BT with SCR and BT with DMTCP. SCR checkpoints are smaller by 62\%-63\%. In addition, there is a steeper increase in DMTCP checkpoints size as the checkpoint size grows from 23.64GiB (1 process) to 26.97GiB (16 processes). On the other hand, the checkpoints size of SCR, which has been developed to scale well, shows a gentle growth in size -- 8.77GiB (1 process) to 9.31GiB (16 processes).


\section{Conclusions and Future Work}
\label{Chapter:Conclusions}
Future exascale supercomputing systems will incorporate non-volatile memories. 
An example for such an HPC system that includes Intel DCPMM, is Aurora \cite{aurora}. 
These new types of non-volatile memories offer simultaneously an ultra-fast storage and a relatively comparable main memory extension, supposedly cheaper than DRAM, with much larger volume per single DIMM. 
By that, these hardware represent a fundamental change, not only to the decades-long memory hierarchy paradigm, but also to high-performance applications and their use-cases. Since scientific HPC applications naturally starve for more byte-addressable memory, as well as for much faster storage, there is a need to understand if and how these applications can optimally utilize non-volatile memory for their benefit. 
This paper investigates whether non-volatile memory can constitute
a much faster replacement for standard storage, 
or a cost-effective alternative for DRAM with regard to high-performance scientific computations.
We carefully study use-cases of the non-volatile memory in current HPC settings, 
and examine representative parallel scientific benchmarks.
We evaluate memory expansion, 
persistent memory storage, 
and explicit/transparent check-point restart configurations. 
We seek to minimize changes to the source codes, 
with the support of an appropriate FS over the NVM. 

Our results show that in all of the suggested tasks, 
DCPMM exhibits outstanding performances. 
Specifically, DCPMM allows to expand the byte-addressable memory 
at the expense of only a few factors of cost; 
it drastically improves access times to persisted storage in comparison 
with SSD, under various file systems; 
and it even facilitated C/R greatly. 

As mentioned in Section~\ref{Chapter:DistributedStorage}, scientific computing relies on DFSs when workloads are executed across multiple nodes. For future work, we plan to set up a multi-node environment (consisting of at least two nodes) and to explore the readiness of DAOS to support Optane™ DCPMM as a distributed storage. Once there are changes in Assise to improve its 
system call interception mechanism and to make it compatible with MPI implementations, we will be able to evaluate Assise and to compare it to DAOS in the context of scientific distributed computing.

We conclude that next generation supercomputing systems might be
able to avoid the use of standard storage completely 
and to adopt non-volatile memories for better performance as well as higher fault tolerance. 
Non-volatile memory should not be overlooked.
Moreover, there are still many possibilities to exploit its full potential 
in future software, specifically, for scientific simulations, 
especially in the presence of new, more advanced technology~\cite{inteloptane}.

\section*{Acknowledgments}
This work was supported by Pazy grant 226/20 and 
the Lynn and William Frankel Center for Computer Science.
Computational support was provided by the NegevHPC project~\cite{negevhpc}. 
The authors would like to thank Israel Hen and Emil Malka for their hardware support.


\bibliographystyle{IEEEtran}
\bibliography{IEEEabrv,biblio_traps_dynamics}

\begin{thebibliography}{10}
\providecommand{\url}[1]{#1}
\csname url@samestyle\endcsname
\providecommand{\newblock}{\relax}
\providecommand{\bibinfo}[2]{#2}
\providecommand{\BIBentrySTDinterwordspacing}{\spaceskip=0pt\relax}
\providecommand{\BIBentryALTinterwordstretchfactor}{4}
\providecommand{\BIBentryALTinterwordspacing}{\spaceskip=\fontdimen2\font plus
\BIBentryALTinterwordstretchfactor\fontdimen3\font minus
  \fontdimen4\font\relax}
\providecommand{\BIBforeignlanguage}[2]{{%
\expandafter\ifx\csname l@#1\endcsname\relax
\typeout{** WARNING: IEEEtran.bst: No hyphenation pattern has been}%
\typeout{** loaded for the language `#1'. Using the pattern for}%
\typeout{** the default language instead.}%
\else
\language=\csname l@#1\endcsname
\fi
#2}}
\providecommand{\BIBdecl}{\relax}
\BIBdecl

\bibitem{liu2021survey}
Liu,~H.-K. \emph{et~al.}, ``A survey of non-volatile main memory technologies:
  State-of-the-arts, practices, and future directions,'' \emph{Journal of
  Computer Science and Technology}, vol.~36, no.~1, pp. 4--32, 2021.

\bibitem{hager2010introduction}
Hager,~G. \emph{et~al.}, \emph{Introduction to high performance computing for
  scientists and engineers}.\hskip 1em plus 0.5em minus 0.4em\relax CRC Press,
  2010.

\bibitem{kaufmann1992supercomputing}
Kaufmann,~W.~J. \emph{et~al.}, \emph{Supercomputing and the Transformation of
  Science}.\hskip 1em plus 0.5em minus 0.4em\relax WH Freeman \& Co., 1992.

\bibitem{snir2014addressing}
Snir,~M. \emph{et~al.}, ``Addressing failures in exascale computing,''
  \emph{The International Journal of High Performance Computing Applications},
  vol.~28, no.~2, pp. 129--173, 2014.

\bibitem{jacob2010memory}
Jacob,~B. \emph{et~al.}, \emph{Memory systems: cache, DRAM, disk}.\hskip 1em
  plus 0.5em minus 0.4em\relax Morgan Kaufmann, 2010.

\bibitem{luttgau2018survey}
L{\"u}ttgau,~J. \emph{et~al.}, ``Survey of storage systems for high-performance
  computing,'' \emph{Supercomputing Frontiers and Innovations}, vol.~5, no.~1,
  pp. 31--58, 2018.

\bibitem{harrod2012journey}
Harrod,~W., ``A journey to exascale computing,'' in \emph{2012 SC Companion:
  High Performance Computing, Networking Storage and Analysis}.\hskip 1em plus
  0.5em minus 0.4em\relax IEEE, 2012, pp. 1702--1730.

\bibitem{fan2001memory}
Fan,~X. \emph{et~al.}, ``Memory controller policies for dram power
  management,'' in \emph{ISLPED'01: Proceedings of the 2001 International
  Symposium on Low Power Electronics and Design (IEEE Cat. No.
  01TH8581)}.\hskip 1em plus 0.5em minus 0.4em\relax IEEE, 2001, pp. 129--134.

\bibitem{oren2016memory}
Oren,~G. \emph{et~al.}, ``Memory-aware management for heterogeneous main memory
  using an optimization of the aging paging algorithm,'' in \emph{2016 45th
  International Conference on Parallel Processing Workshops (ICPPW)}.\hskip 1em
  plus 0.5em minus 0.4em\relax IEEE, 2016, pp. 98--105.

\bibitem{bergman2008exascale}
Bergman,~K. \emph{et~al.}, ``Exascale computing study: Technology challenges in
  achieving exascale systems,'' \emph{Defense Advanced Research Projects Agency
  Information Processing Techniques Office (DARPA IPTO), Tech. Rep}, vol.~15,
  2008.

\bibitem{moody2010design}
Moody,~A. \emph{et~al.}, ``Design, modeling, and evaluation of a scalable
  multi-level checkpointing system,'' in \emph{SC'10: Proceedings of the 2010
  ACM/IEEE International Conference for High Performance Computing, Networking,
  Storage and Analysis}.\hskip 1em plus 0.5em minus 0.4em\relax IEEE, 2010, pp.
  1--11.

\bibitem{wang2010hybrid}
Wang,~C. \emph{et~al.}, ``Hybrid checkpointing for mpi jobs in hpc
  environments,'' in \emph{2010 IEEE 16th International Conference on Parallel
  and Distributed Systems}.\hskip 1em plus 0.5em minus 0.4em\relax IEEE, 2010,
  pp. 524--533.

\bibitem{weiland2019early}
Weiland,~M. \emph{et~al.}, ``An early evaluation of intel's optane dc
  persistent memory module and its impact on high-performance scientific
  applications,'' in \emph{Proceedings of the International Conference for High
  Performance Computing, Networking, Storage and Analysis}, 2019, pp. 1--19.

\bibitem{wu2020lessons}
Wu,~Y. \emph{et~al.}, ``Lessons learned from the early performance evaluation
  of intel optane dc persistent memory in dbms,'' in \emph{Proceedings of the
  16th International Workshop on Data Management on New Hardware}, 2020, pp.
  1--3.

\bibitem{weiland2020usage}
Weiland,~M. \emph{et~al.}, ``Usage scenarios for byte-addressable persistent
  memory inhigh-performance and data intensive computing,'' \emph{arXiv
  preprint arXiv:2012.06473}, 2020.

\bibitem{peng2020demystifying}
Peng,~I. \emph{et~al.}, ``Demystifying the performance of hpc scientific
  applications on nvm-based memory systems,'' in \emph{2020 IEEE International
  Parallel and Distributed Processing Symposium (IPDPS)}.\hskip 1em plus 0.5em
  minus 0.4em\relax IEEE, 2020, pp. 916--925.

\bibitem{wu2021runtime}
Wu,~K., ``Runtime data management on non-volatile memory-based high performance
  systems,'' Ph.D. dissertation, University of California, Merced, 2021.

\bibitem{christgau2020leveraging}
Christgau,~S. \emph{et~al.}, ``Leveraging a heterogeneous memory system for a
  legacy fortran code: The interplay of storage class memory, dram and os,'' in
  \emph{2020 IEEE/ACM Workshop on Memory Centric High Performance Computing
  (MCHPC)}.\hskip 1em plus 0.5em minus 0.4em\relax IEEE, 2020, pp. 17--24.

\bibitem{ren2021optimizing}
Ren,~J. \emph{et~al.}, ``Optimizing large-scale plasma simulations on
  persistent memory-based heterogeneous memory with effective data placement
  across memory hierarchy,'' in \emph{Proceedings of the ACM International
  Conference on Supercomputing}, 2021, pp. 203--214.

\bibitem{malinowski2019multi}
Malinowski,~A. \emph{et~al.}, ``Multi-agent large-scale parallel crowd
  simulation with nvram-based distributed cache,'' \emph{Journal of
  Computational Science}, vol.~33, pp. 83--94, 2019.

\bibitem{garg2020need}
Garg,~S. \emph{et~al.}, ``The need for precise and efficient memory capacity
  budgeting,'' in \emph{The International Symposium on Memory Systems}, 2020,
  pp. 169--177.

\bibitem{zhu2021empirical}
Zhu,~G. \emph{et~al.}, ``An empirical evaluation of nvm-aware file systems on
  intel optane dc persistent memory modules,'' in \emph{2021 International
  Conference on Information Networking (ICOIN)}.\hskip 1em plus 0.5em minus
  0.4em\relax IEEE, 2021, pp. 559--564.

\bibitem{bother2021drop}
B{\"o}ther,~M. \emph{et~al.}, ``Drop it in like it’s hot: An analysis of
  persistent memory as a drop-in replacement for nvme ssds,'' in
  \emph{Proceedings of the 17th International Workshop on Data Management on
  New Hardware (DaMoN 2021)}, 2021, pp. 1--8.

\bibitem{soumagne2021accelerating}
Soumagne,~J. \emph{et~al.}, ``Accelerating hdf5 i/o for exascale using daos,''
  \emph{IEEE Transactions on Parallel and Distributed Systems}, 2021.

\bibitem{lopez2021exploring}
L{\'o}pez-G{\'o}mez,~J. \emph{et~al.}, ``Exploring object stores for
  high-energy physics data storage,'' \emph{arXiv preprint arXiv:2107.07304},
  2021.

\bibitem{wong2003parallel}
Wong,~P. \emph{et~al.}, ``Nas parallel benchmarks i/o version 2.4,'' \emph{NASA
  Ames Research Center, Moffet Field, CA, Tech. Rep. NAS-03-002}, 2003.

\bibitem{ren2019easycrash}
Ren,~J. \emph{et~al.}, ``Easycrash: Exploring non-volatility of non-volatile
  memory for high performance computing under failures,'' \emph{arXiv preprint
  arXiv:1906.10081}, 2019.

\bibitem{zvi2021optimized}
Zvi,~K. \emph{et~al.}, ``Optimized memoryless fair-share hpc resources
  scheduling using transparent checkpoint-restart preemption,'' \emph{arXiv
  preprint arXiv:2102.12953}, 2021.

\bibitem{ren2020exploring}
Ren,~J. \emph{et~al.}, ``Exploring non-volatility of non-volatile memory for
  high performance computing under failures,'' in \emph{2020 IEEE International
  Conference on Cluster Computing (CLUSTER)}.\hskip 1em plus 0.5em minus
  0.4em\relax IEEE, 2020, pp. 237--247.

\bibitem{mcanlis1984data}
McAnlis,~J.~C. \emph{et~al.}, ``Data processor system including data-save
  controller for protection against loss of volatile memory information during
  power failure,'' Jul.~3 1984, uS Patent 4,458,307.

\bibitem{peng2019system}
Peng,~I.~B. \emph{et~al.}, ``System evaluation of the intel optane
  byte-addressable nvm,'' in \emph{Proceedings of the International Symposium
  on Memory Systems}, 2019, pp. 304--315.

\bibitem{Scargall2020}
\BIBentryALTinterwordspacing
Scargall,~S., \emph{Introducing the Persistent Memory Development Kit}.\hskip
  1em plus 0.5em minus 0.4em\relax Berkeley, CA: Apress, 2020, pp. 63--72.
  [Online]. Available: \url{https://doi.org/10.1007/978-1-4842-4932-1_5}
\BIBentrySTDinterwordspacing

\bibitem{axboe2014fio}
Axboe,~J., ``Fio-flexible io tester,''
  \url{http://freshmeat.sourceforge.net/projects/fio}, 2014.

\bibitem{izraelevitz2019basic}
Izraelevitz,~J. \emph{et~al.}, ``Basic performance measurements of the intel
  optane dc persistent memory module,'' \emph{arXiv preprint arXiv:1903.05714},
  2019.

\bibitem{bailey1991parallel}
Bailey,~D.~H. \emph{et~al.}, ``The nas parallel benchmarks,'' \emph{The
  International Journal of Supercomputing Applications}, vol.~5, no.~3, pp.
  63--73, 1991.

\bibitem{polybench}
Pouchet,~L.-N. \emph{et~al.}, ``{PolyBench Benchmarks},''
  \url{https://web.cse.ohio-state.edu/~pouchet.2/software/polybench/},
  [Online].

\bibitem{bailey2011parallel}
Bailey,~D.~H., ``Nas parallel benchmarks,'' in \emph{Encyclopedia of Parallel
  Computing}.\hskip 1em plus 0.5em minus 0.4em\relax Springer, 2011, pp.
  1254--1259.

\bibitem{yuki2014understanding}
Yuki,~T., ``Understanding polybench/c 3.2 kernels,'' in \emph{International
  workshop on Polyhedral Compilation Techniques (IMPACT)}, 2014, pp. 1--5.

\bibitem{gabriel2004open}
Gabriel,~E. \emph{et~al.}, ``Open mpi: Goals, concept, and design of a next
  generation mpi implementation,'' in \emph{European Parallel Virtual
  Machine/Message Passing Interface Users’ Group Meeting}.\hskip 1em plus
  0.5em minus 0.4em\relax Springer, 2004, pp. 97--104.

\bibitem{tristian2019analyzing}
Tristian,~T. \emph{et~al.}, ``Analyzing the performance of intel optane dc
  persistent memory in app direct mode in lenovo thinksystem servers,'' 2019.

\bibitem{intel-quick-start-guide-persistent-memory}
``{Quick Start Guide Part 1: Persistent Memory Provisioning Introduction},''
  \url{https://software.intel.com/content/www/us/en/develop/articles/qsg-intro-to-provisioning-pmem.html},
  [Online].

\bibitem{nova}
\BIBentryALTinterwordspacing
Xu,~J. \emph{et~al.}, ``{NOVA}: A log-structured file system for hybrid
  volatile/non-volatile main memories,'' in \emph{14th {USENIX} Conference on
  File and Storage Technologies ({FAST} 16)}.\hskip 1em plus 0.5em minus
  0.4em\relax Santa Clara, CA: {USENIX} Association, Feb. 2016, pp. 323--338.
  [Online]. Available:
  \url{https://www.usenix.org/conference/fast16/technical-sessions/presentation/xu}
\BIBentrySTDinterwordspacing

\bibitem{KadekodiEtAl19-SplitFS}
Kadekodi,~R. \emph{et~al.}, ``{SplitFS: Reducing Software Overhead in File
  Systems for Persistent Memory},'' in \emph{Proceedings of the 27th ACM
  Symposium on Operating Systems Principles (SOSP '19)}, Ontario, Canada,
  October 2019.

\bibitem{octopus}
\BIBentryALTinterwordspacing
Lu,~Y. \emph{et~al.}, ``Octopus: an rdma-enabled distributed persistent memory
  file system,'' in \emph{2017 {USENIX} Annual Technical Conference ({USENIX}
  {ATC} 17)}.\hskip 1em plus 0.5em minus 0.4em\relax Santa Clara, CA: {USENIX}
  Association, Jul. 2017, pp. 773--785. [Online]. Available:
  \url{https://www.usenix.org/conference/atc17/technical-sessions/presentation/lu}
\BIBentrySTDinterwordspacing

\bibitem{assise}
\BIBentryALTinterwordspacing
Anderson,~T.~E. \emph{et~al.}, ``Assise: Performance and availability via
  client-local {NVM} in a distributed file system,'' in \emph{14th {USENIX}
  Symposium on Operating Systems Design and Implementation ({OSDI} 20)}.\hskip
  1em plus 0.5em minus 0.4em\relax {USENIX} Association, Nov. 2020, pp.
  1011--1027. [Online]. Available:
  \url{https://www.usenix.org/conference/osdi20/presentation/anderson}
\BIBentrySTDinterwordspacing

\bibitem{daos-springer}
Liang,~Z. \emph{et~al.}, ``Daos: A scale-out high performance storage stack for
  storage class memory,'' in \emph{Supercomputing Frontiers}, Panda,~D.~K.,
  Ed.\hskip 1em plus 0.5em minus 0.4em\relax Cham: Springer International
  Publishing, 2020, pp. 40--54.

\bibitem{daos-website}
``{DAOS official documentation website},'' \url{https://daos-stack.github.io/},
  [Online].

\bibitem{strata}
\BIBentryALTinterwordspacing
Kwon,~Y. \emph{et~al.}, ``Strata: A cross media file system,'' ser. SOSP
  '17.\hskip 1em plus 0.5em minus 0.4em\relax New York, NY, USA: Association
  for Computing Machinery, 2017, p. 460–477. [Online]. Available:
  \url{https://doi.org/10.1145/3132747.3132770}
\BIBentrySTDinterwordspacing

\bibitem{assise-github-issue-7}
``{Assise's Github issue \#7},''
  \url{https://github.com/ut-osa/assise/issues/7}, [Online].

\bibitem{assise-github-issue-3-comment}
``{Assise's Github issue \#3},''
  \url{https://github.com/ut-osa/assise/issues/3#issuecomment-838785034},
  [Online].

\bibitem{daos-website-posix}
``{POSIX Namespace - DAOS v1.2},''
  \url{https://daos-stack.github.io/user/posix/}, [Online].

\bibitem{scrGit}
``{Scalable Checkpoint / Restart (SCR) Library github page.}''
  \url{https://github.com/LLNL/scr}, [Online].

\bibitem{scrDocs}
``{Scalable Checkpoint / Restart (SCR) Library documentation page.}''
  \url{https://scr.readthedocs.io/en/latest/}, [Online].

\bibitem{ansel2009dmtcp}
Ansel,~J. \emph{et~al.}, ``{DMTCP}: Transparent checkpointing for cluster
  computations and the desktop,'' in \emph{2009 IEEE International Symposium on
  Parallel \& Distributed Processing (IPDPS'09)}.\hskip 1em plus 0.5em minus
  0.4em\relax Rome, Italy: IEEE, 2009, pp. 1--12.

\bibitem{rodriguez2019job}
Rodr{\'\i}guez-Pascual,~M. \emph{et~al.}, ``Job migration in hpc clusters by
  means of checkpoint/restart,'' \emph{The Journal of Supercomputing}, vol.~75,
  no.~10, pp. 6517--6541, 2019.

\bibitem{cao_transparent_2013}
Cao,~J. \emph{et~al.}, ``Transparent {Checkpoint}-{Restart} over
  {InfiniBand},'' \emph{HPDC 2014 - Proceedings of the 23rd International
  Symposium on High-Performance Parallel and Distributed Computing}, Dec. 2013.

\bibitem{maronas2020extending}
Maro{\~n}as,~M. \emph{et~al.}, ``Extending the openchk model with advanced
  checkpoint features,'' \emph{Future Generation Computer Systems}, vol. 112,
  pp. 738--750, 2020.

\bibitem{aurora}
``{Aurora | Argonne Leadership Computing Facility},''
  \url{https://alcf.anl.gov/aurora}, 2021, [Online].

\bibitem{inteloptane}
``{Intel® Optane™ persistent memory (PMem)},''
  \url{https://www.intel.com/content/www/us/en/products/details/memory-storage/optane-dc-persistent-memory.html},
  [Online].

\bibitem{negevhpc}
``{NegevHPC Project},'' \url{https://www.negevhpc.com}, [Online].

\end{thebibliography}

\end{document}